\documentclass[pdflatex,sn-mathphys-num]{sn-jnl}

\usepackage{graphicx}%
\usepackage{multirow}%
\usepackage{amsmath,amssymb,amsfonts}%
\usepackage{amsthm}%
\usepackage{mathrsfs}%
\usepackage[title]{appendix}%
\usepackage{xcolor}%
\usepackage{textcomp}%
\usepackage{url}

\usepackage{manyfoot}%
\usepackage{booktabs}%
\usepackage{algorithm}%
\usepackage{algorithmicx}%
\usepackage{algpseudocode}%
\usepackage{listings}%
\usepackage{pdflscape}
\usepackage{geometry}
\usepackage{lineno}

\theoremstyle{thmstyleone}%
\theoremstyle{thmstyletwo}%

\theoremstyle{thmstylethree}%

\begin{document}

\title[Magnetic rigidity in SS~433 jets]{Magnetic rigidity reveals the PeVatron acceleration region in SS~433}



\author[1,2]{\fnm{Jie} \sur{Liao}}\email{liaoj68@mail2.sysu.edu.cn}
\equalcont{These authors contributed equally to this work.}

\author[2,3,4]{\fnm{Wancheng} \sur{Xu}}\email{xuwancheng@xao.ac.cn}
\equalcont{These authors contributed equally to this work.}

\author*[2,3,5]{\fnm{Lang} \sur{Cui}}\email{cuilang@xao.ac.cn}
\author*[1]{\fnm{Pak-Hin Thomas} \sur{Tam}}\email{tanbxuan@mail.sysu.edu.cn}

\author[2]{\fnm{Xi} \sur{Yan}}\email{yanxi@xao.ac.cn}
\author[1]{\fnm{Linuo} \sur{Yang}}\email{yangln26@mail2.sysu.edu.cn}
\author[6,7]{\fnm{Ruo-Yu} \sur{Liu}}\email{ryliu@nju.edu.cn}
\author[8]{\fnm{Liang} \sur{Chen}}\email{chenliang@shao.ac.cn}
\author[2]{\fnm{Ning} \sur{Chang}}\email{changning@xao.ac.cn}
\author[6,7]{\fnm{Yong-Feng} \sur{Huang}}\email{hyf@nju.edu.cn}
\author[1]{\fnm{Long} \sur{Ji}}\email{jilong@mail.sysu.edu.cn}
\author[2]{\fnm{Ashutosh} \sur{Tripathi}}\email{ashutoshtripathi@xao.ac.cn}
\author[9,10]{\fnm{Felix} \sur{Aharonian}}\email{Felix.Aharonian@mpi-hd.mpg.de}

\affil[1]{\orgdiv{School of Physics and Astronomy}, 
\orgname{Sun Yat-sen University}, 
\orgaddress{\street{No. 2 Daxue Road, Xiangzhou District}, \city{Zhuhai}, \postcode{519082}, \state{Guangdong}, \country{China}}}

\affil[2]{\orgdiv{State Key Laboratory of Radio Astronomy and Technology}, \orgname{Xinjiang Astronomical Observatory, CAS}, \orgaddress{\street{150 Science 1-Street}, \city{Urumqi}, \postcode{830011}, \state{Xinjiang}, \country{China}}}

\affil[3]{\orgdiv{School of Astronomy and Space Science}, \orgname{ University of Chinese Academy of Sciences}, \orgaddress{\street{No. 1 Yanqihu East Road}, \city{Beijing}, \postcode{101408}, \country{China}}}

\affil[4]{\orgdiv{Konkoly Observatory}, \orgname{HUN-REN Research Centre for Astronomy and Earth Sciences, MTA Centre of Excellence}, \orgaddress{\street{Konkoly Thege Mikl\'os \'ut 15-17}, \city{Budapest}, \postcode{H-1121}, \country{Hungary}}}

\affil[5]{\orgdiv{Xinjiang Key Laboratory of Radio Astrophysics}, \orgname{Xinjiang Astronomical Observatory, CAS}, \orgaddress{\street{150 Science 1-Street}, \city{Urumqi}, \postcode{830011}, \state{Xinjiang}, \country{China}}}

\affil[6]{\orgdiv{School of Astronomy and Space Science}, \orgname{ Nanjing University}, \orgaddress{\street{163 Xianlin Avenue}, \city{Nanjing}, \postcode{210023}, \state{Jiangsu}, \country{China}}}
\affil[7]{\orgdiv{Key Laboratory of Modern Astronomy and Astrophysics (Nanjing University)}, \orgname{Ministry of Education}, \orgaddress{\street{163 Xianlin Avenue}, \city{Nanjing}, \postcode{210023}, \state{Jiangsu}, \country{China}}}
\affil[8]{\orgdiv{State Key Laboratory of Radio Astronomy and Technology}, \orgname{Shanghai Astronomical Observatory, CAS}, \orgaddress{\street{80 Nandan Road}, \city{Shanghai}, \postcode{200030},  \country{China}}}

\affil[9]{%
\orgname{Yerevan State University},
\orgaddress{%
\street{1 Alek Manukyan Street},
\city{Yerevan},
\postcode{0025},
\country{Armenia}}}
\affil[10]{%
\orgname{Max-Planck-Institut for Nuclear Physics},
\orgaddress{%
\street{P.O. Box 103980},
\city{Heidelberg},
\postcode{69029},
\country{Germany}}}


\abstract{
\textbf{PeVatrons are cosmic accelerators capable of driving particles to petaelectronvolt (PeV) energies.
Recently, microquasar jets have emerged as compelling Galactic PeVatron candidates. This is especially the case for SS~433 as its $>$100~TeV gamma-ray emission is spatially coincident with an atomic cloud. However, the exact region where PeV protons are accelerated and injected within these jets remains unresolved. 
Here we report, using archival, multi-frequency VLBA observations, the magnetic field profile $B(H)$ along the SS~433 inner jet on tens of AU scale, where $H$ is the distance from the central compact object. We find that the field declines as $B(H)\propto H^{-0.50\pm0.12}$, demonstrating that the magnetic rigidity $B(H) R_{\rm acc}$ grows with $H$ for a conical jet.
This implies the Hillas limit ($E_{\rm max}\propto B H$) to lie well beyond a PeV at a few hundred-AU scale, which becomes a highly potential site for accelerating protons to energies $E_{\rm cut}\simeq2.6\,\mathrm{PeV}$ inferred from the LHAASO gamma-ray spectrum. These results reveal a hidden PeVatron within the baryonic ejecta of microquasar SS~433, well upstream of the extended TeV-emitting lobes.
}
}

\maketitle

\section{Introduction}\label{sec1}

Identifying PeVatrons, i.e., Galactic accelerators that reach cosmic-ray knee energies, i.e., 10$^{15}$--10$^{16}$~eV, remains a central problem in high-energy astrophysics   \citep{Blasi2013,Gaisser2016}.
Supernova remnants are the prominent acceleration sites for Galactic cosmic-ray production \citep{Bell1978,BlandfordOstriker1978,Drury1983,BlandfordEichler1987}, but gamma-ray observations above tens of tera-electron volts show that alternative sites, other than supernova remnants, are required to explain the cosmic-ray spectrum up to the knee \citep{Abramowski2016GCPeVatron,Albert2021LHAASO}.
Among the emerging PeVatron candidates, relativistic jets from accreting compact
objects (``microquasars'') are particularly promising. They combine large
kinetic power with magnetized, mildly relativistic outflows. Shocks,
shear and magnetic dissipation within these jets, together with terminal interactions with the surrounding medium, are among possible ways to accelerate particle\citep{Peretti2025, Wang2025, Wan2026}.

SS~433 provides a natural laboratory to study this problem.
It is a persistent, supercritically accreting X-ray binary embedded in the W50 nebula \citep{Fabrika2004}, launching baryonic jets with a bulk speed of $\simeq 0.26c$ \citep{Margon1984,Fabrika2004}.
These jets are well resolved on length scales that differ by orders of magnitude -- from milliarcsecond scales out to the large-scale W50 lobes \citep{Paragi1999}. 
High-energy observations have established the system as an efficient particle accelerator. The first resolved TeV emission from the eastern and western jet-cloud interaction regions \citep{Abeysekara2018SS433} was detected by the High-Altitude Water Cherenkov (HAWC) Observatory. The High Energy Stereoscopic System (H.E.S.S.) subsequently mapped energy-dependent TeV structures along the jets which clearly link the lower-energy TeV component to downstream shock regions on parsec scales \citep{HESS2024SS433}.
More recently, Large High Altitude Air Shower Observatory (LHAASO) detected ultra-high-energy ($>$100~TeV) gamma-ray emission from the SS~433/W50 region, which requires the parent-particle energies to be on the order of PeV~
\citep{LHAASO2025Microquasars}.

\begin{figure}[h]
\centering
\includegraphics[width=0.9\textwidth]{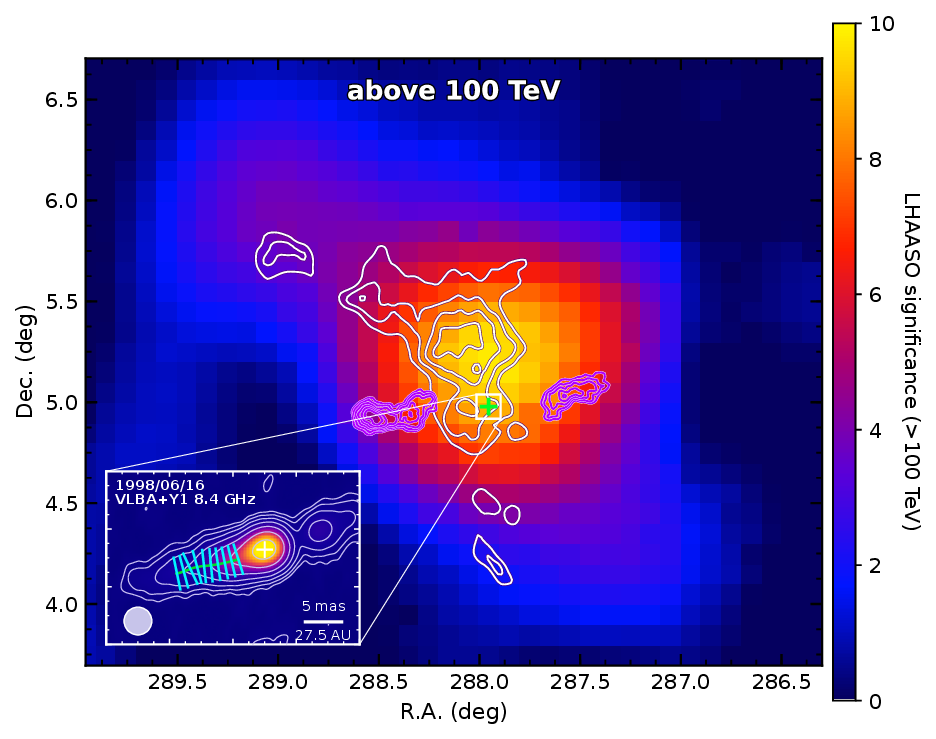}
\caption{\textbf{Multi-wavelength view of the SS~433 field.} The background image shows the released LHAASO $>100\,\mathrm{TeV}$
significance map in equatorial coordinates \citep{LHAASO2025Microquasars}, with the colour bar giving the LHAASO significance. Purple contours show the H.E.S.S. significance map \citep{HESS2024SS433} after WCS transformation onto the LHAASO image frame, and white contours trace the H\,{\sc i} gas distribution used as the neutral-target reference \citep{HI4PI2016}. The green cross marks the radio position of SS~433, and the small white box marks the region enlarged in the inset. The inset shows the 1998 June 16 VLBA+Y1 8.4\,GHz image restored from the radio FITS map; grey contours are drawn at $3\sigma\times2^n$ with $\sigma=0.43\,\mathrm{mJy\,beam^{-1}}$. The green curve follows the radio ridge line, cyan bars mark the transverse slices used for the resolved jet analysis, the white cross marks the central binary position, the filled circle shows the restoring beam, and the white bar gives the 5\,mas scale corresponding to 27.5\,AU. Thin white connector lines link the SS~433 zoom box to the VLBA inset.}\label{fig:hess_lhaaso_overlay}
\end{figure}

The LHAASO $>$100~TeV gamma-ray observation has established SS~433 as a promising microquasar PeVatron candidate, but it does not indicate where the PeV particles are first accelerated. 
The $>$100~TeV gamma-ray emission is concentrated around the central SS~433/W50 region and exhibits a quasi-isotropic morphology (Figure~\ref{fig:hess_lhaaso_overlay}), which is in marked contrast to the bipolar~
morphology observed at TeV energies \citep{Abeysekara2018SS433,HESS2024SS433,LHAASO2025Microquasars}. 
Such morphological difference suggests that the $>$100~TeV component is not simply a high-energy continuation of the downstream lobe emission, but instead traces a population of particles, whose injection sites differ substantially from the latter. 
Current gamma-ray instruments, however, cannot resolve the AU-scale radio jet where such acceleration and injection could begin. 
Locating the accelerator, therefore, requires an independent probe of the physical conditions in the inner baryonic ejecta. 
Very Long Baseline Array (VLBA) provides this probe by resolving the jet on $\sim10^{-4}$$\,\mathrm{pc}$ scales \citep{Paragi1999}. The spatially resolved local synchrotron luminosity can be measured, which, under minimum-energy assumptions, can be used to infer the magnetic field strength along the direction of the jet.



For a proton to reach PeV energies, its Larmor radius must be smaller than the transverse extent of the accelerator. For a jet moving at speed $\beta_j c$, the corresponding speed-weighted Hillas limit is $E_{\mathrm{H},\beta}\simeq \beta_j ZeBR_{\mathrm{acc}}$ \citep{Hillas1984,RiegerBoschRamonDuffy2007,KoteraOlinto2011}. For a conical jet, $R_{\mathrm{acc}}\simeq\alpha_j H$, the magnetic rigidity scales as $B(H)H$ and therefore increases downstream when the magnetic field declines more slowly than $H^{-1}$. 

Here we construct a spatially resolved profile of $B(H)$ along the baryonic ejecta of SS~433 with multi-frequency VLBA observations. We find $B(H)\propto H^{-0.50\pm0.12}$, which, under the fiducial geometry, brings the speed-weighted Hillas energy into the few-PeV range on hundred-AU scales, consistent with the maximal proton energy required by the emission above $100 \,\mathrm{TeV}$ from LHAASO. These results reveal a highly potential compact PeV-proton acceleration and injection region within the radio-resolved ejecta.

\section{Results}
\label{sec2}

We construct the magnetic-rigidity profile along the radio-resolved ejecta of SS~433. 
Archival VLBA+Y1 (where Y1 denotes a single element of the VLA) observations at 5, 8.4 and 15\,GHz are restored to a common circular beam of $\Theta_{\rm maj}=3.537$\,mas and are registered to a common frame using the frequency-dependent core positions of \citet{Yan2026SS433Collimation}, allowing the eastern jet to be sliced transversely at matched resolution. 
Ten slices spanning $3.6$--$11.5$\,mas ($24$--$75$\,AU deprojected for $\theta_{\rm view}=57^{\circ}$ and $D=5.5$\,kpc \citep{BlundellBowler2004}) satisfy our quality criteria, and for each we measured the local flux density at all three frequencies, the beam-deconvolved transverse width, and the two-point spectral indices. 
The spectra show an overall outward steepening, with $\alpha_{8.4}^{15}$ ($S_{\nu}\propto\nu^{-\alpha}$) rising from $0.35\pm0.03$ to $0.86\pm0.25$, confirming that the slices sample optically thin synchrotron emission and allowing the Pacholczyk constant to be evaluated slice by slice.

Under the standard minimum-energy assumption \citep{Pacholczyk1970}, these observables yield an equipartition field for each slice, with uncertainties of each slice propagated by Monte Carlo over the flux densities, widths, and spectral indices. The absolute normalization carries a systematic uncertainty from the constrained non-radiating particle content, filling factor, and transverse geometry. We therefore treat the equipartition value as a 
commonly used reference scale rather than a direct field measurement. For the case $k=1$ the resolved ejecta have fields of $0.20$--$0.35\,\mathrm{G}$ on the scales of tens of AU. Varying $k$ from $0$ to $100$ changes the normalization by the factor of four but leaves the radial dependence unchanged, since $k$ enters only as a multiplicative constant (see Methods~\ref{Methods} for details).

\begin{figure}[h]
\centering
\includegraphics[width=0.9\textwidth]{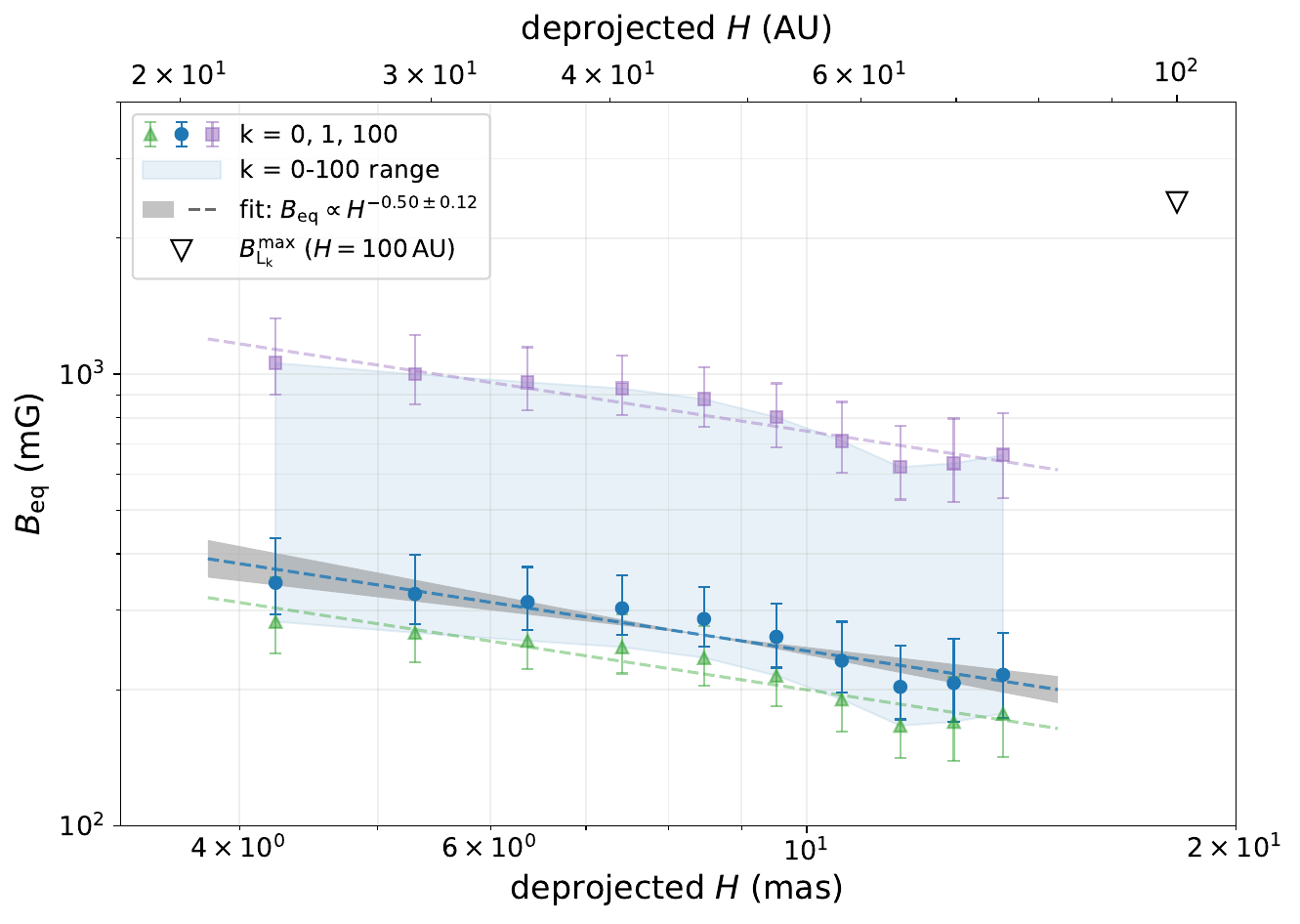}
\caption{\textbf{Equipartition magnetic field along the eastern jet of SS~433.} The ten retained slices are plotted as a function of deprojected distance H, with error bars showing the Monte-Carlo 1$\sigma$ uncertainties. Green triangles, blue circles, and purple squares show $k=0$, $1$, and $100$, respectively, where $k$ is the ratio of non-radiating to radiating particle energy; the shaded band spans the full $k=0$--$100$ range. The dashed lines are the least-squares power-law fit, $B_{\rm eq}\propto H^{-0.50\pm0.12}$. The grey wedge shows this slope uncertainty, pivoted at the mean $\log H$ of the fitted slices. The inverted triangle at $H = 100$~AU (top-right) marks $B^{\rm max}_{L_{\rm k}}\sim2.5$~G, the upper limit obtained by requiring the local internal energy density to not exceed the jet kinetic luminosity $L_{\rm k}$.}\label{fig:1998_B_eq}
\end{figure}

Figure~\ref{fig:1998_B_eq} shows that the magnetic field declines along the compact ejecta as
$B(H)\propto H^{-0.50\pm0.12}$, with the uncertainty dominated by coherent calibration and width systematics.
Under the fiducial prescription $R_{\rm acc}\simeq\alpha_jH$, this profile gives
$E_{\rm H,\beta}\propto B(H)H\propto H^{0.50\pm0.12}$, so that the speed-weighted Hillas energy rises downstream.
For $\alpha_j\simeq0.1$ and $\beta_j\simeq0.26$, the fitted profile yields
$E_{\rm H,\beta}\sim2\,\mathrm{PeV}$ at $H\simeq100\,\mathrm{AU}$ and reaches the few-PeV range on hundred-AU scales, consistent with the parent-proton energies inferred from the LHAASO emission above $100\,\mathrm{TeV}$.
This conclusion is insensitive to the adopted transverse geometry.
Using the measured jet width, $R_{\rm acc}=W/2$, instead of $\alpha_jH$ changes the numerical Hillas energy and nominal crossing distance, but still supports PeV-scale confinement. Moreover, at fixed jet power and magnetization, the Poynting-limited product $BR_{\rm acc}$ is independent of the transverse scale.

\begin{figure}[t]
\centering
\includegraphics[width=1.0\textwidth]{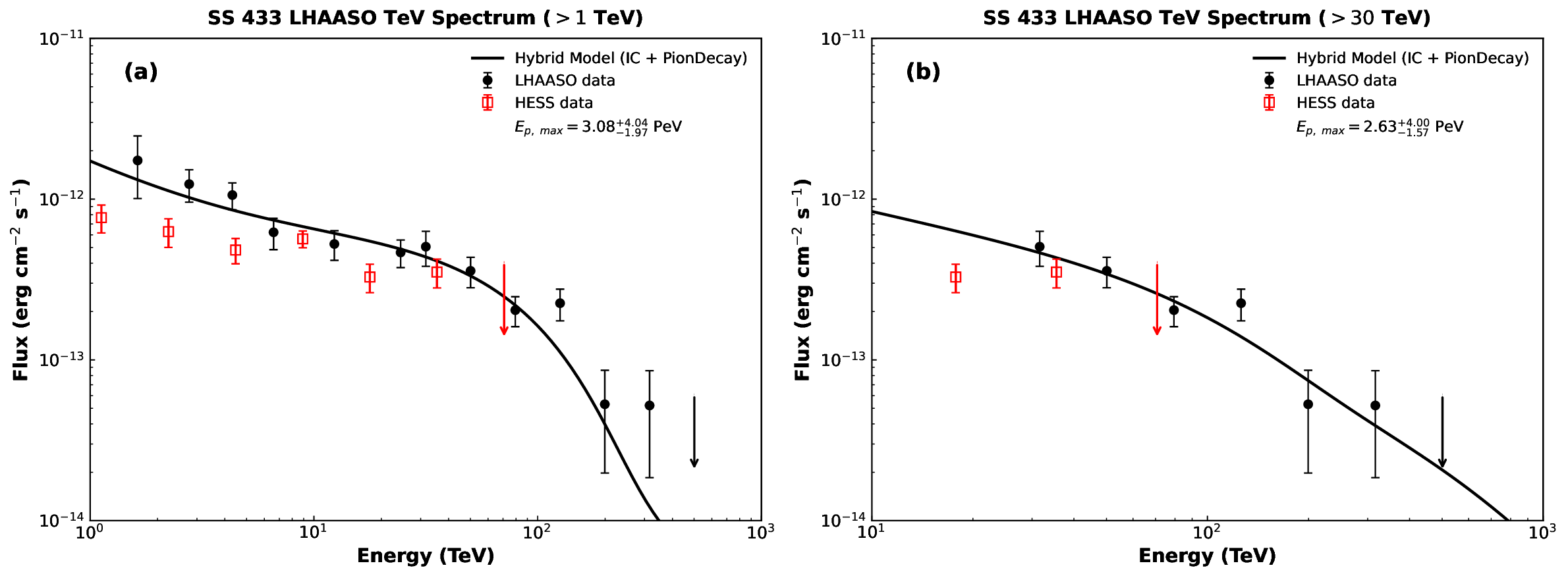}
\caption{
\textbf{Lepto-hadronic fit to the LHAASO spectrum of SS~433.} \citep{LHAASO2025Microquasars}. It represents the fit results for two energy ranges ($>1\,\mathrm{TeV}$ and $>30\,\mathrm{TeV}$).
The fit to the $>1\,\mathrm{TeV}$ data yields $E_{\rm p,max}=3.08^{+4.04}_{-1.97}\,\mathrm{PeV}$, while the fit to the $>30\,\mathrm{TeV}$ data gives $E_{\rm p,max}=2.63^{+4.00}_{-1.57}\,\mathrm{PeV}$.
H.E.S.S.\ measurements \citep{HESS2024SS433} are shown in red for comparison and are excluded from the fit.
}
\label{fig:LHAAS0_hybrid_sed}
\end{figure}

The TeV--PeV emission detected by H.E.S.S. and LHAASO extends far beyond the AU-scale jet, with LHAASO resolving an extended central component above \(100\,\mathrm{TeV}\) \citep{HESS2024SS433,LHAASO2025Microquasars}.
Producing such $>$100~TeV gamma rays through inverse-Compton emission only inside the compact ejecta would require PeV electrons, which would cool too rapidly in such strong magnetic fields \citep{RybickiLightman1979, BlumenthalGould1970}. Such rapid synchrotron cooling may also contribute to the X-ray gap between the compact ejecta and the parsec-scale jets, where non-thermal X-ray emission reappears at $\sim25$~pc from the binary \citep{HESS2024SS433}.
In a lepto-hadronic fit, by contrast, inverse-Compton emission contributes at lower TeV energies and pion-decay emission dominates at the highest energies. This lepto-hadronic fit yields a parent-proton cutoff of $E_{p,\rm cut}=2.63^{+4.00}_{-1.57}\,\mathrm{PeV}$ (Figure~\ref{fig:LHAAS0_hybrid_sed}). This highest energy is comfortably reached by the Hillas estimate at $H\simeq140\,\mathrm{AU}$.

We next examine whether acceleration can proceed sufficiently fast within the resolved ejecta.
At $H\simeq100\,\mathrm{AU}$, the near-Bohm first-order Fermi acceleration to a few PeV is faster than the dynamical flow time, $t_{\rm dyn}\simeq2.2\,\mathrm{d}$.
Moreover, at fixed proton energy, $t_{\rm acc}\propto B^{-1}\propto H^{0.50\pm0.12}$ whereas $t_{\rm dyn}\propto H$, so that $t_{\rm acc}/t_{\rm dyn}$ decreases downstream.
The acceleration condition therefore becomes progressively less restrictive with distance.
The main residual uncertainty lies in the particle residence time, such as free transverse diffusion places the most conservative confinement limit, whereas advection and longitudinal trapping can extend the effective residence time.
Within this range of transport geometries, the compact ejecta remain compatible with the acceleration and injection of PeV protons.

The radio-derived equipartition field profile already indicates that the compact ejecta possess sufficient magnetic rigidity to accelerate PeV protons.
The conclusion is not restricted to the fiducial minimum-energy condition. Stronger fields permitted by the jet-power budget, potentially reaching a few gauss, would simply shift the acceleration scale to smaller acceleration radii.
However, the acceleration site would not necessarily coincide with the gamma-ray radiation zone.
Because the $pp$ cooling time within the inner jet greatly exceeds the local flow time \citep{Kelner2006,Kafexhiu2014}, only a negligible fraction of the accelerated-proton energy can be radiated inside the compact ejecta.
The PeV protons can instead escape or be advected into the surrounding W50 environment, where interactions with ambient gas produce extended $>$100~TeV emission.
The LHAASO $>100\,\mathrm{TeV}$ emission centralised toward the central region of SS~433 and its overlap with the neutral $\mathrm{H\,I}$ gas are consistent with this picture of a compact PeV accelerator feeding a larger hadronic reservoir.

\begin{figure}[t]
\centering
\includegraphics[width=0.7\textwidth]{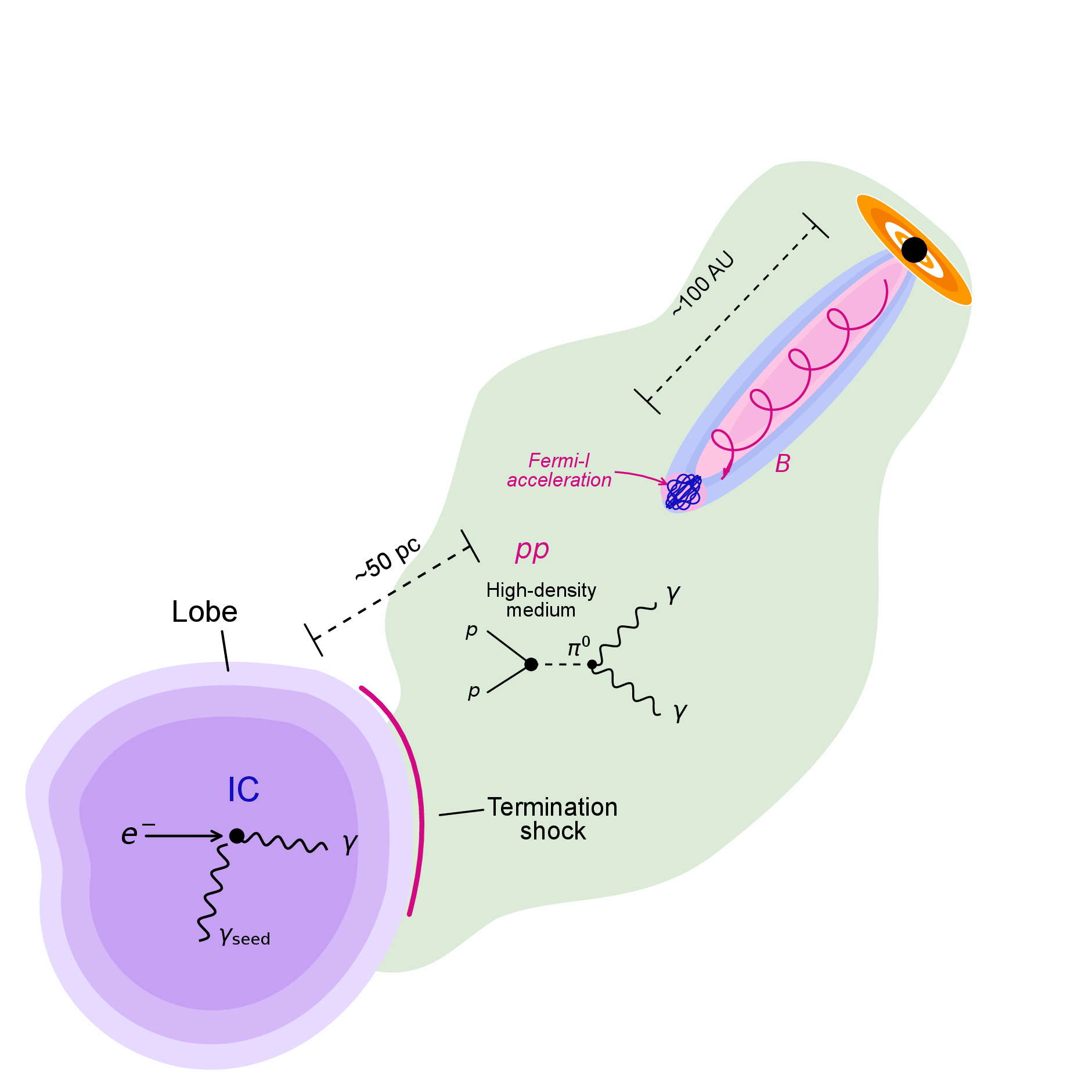}
\caption{
\textbf{Multi-zone PeVatron picture for SS~433. }
The compact radio ejecta on tens-to-hundreds of AU scales provide a viable PeV-proton injection channel. 
After transport into the surrounding W50 environment, these protons interact with neutral gas and produce the observed $>$100~TeV gamma rays through $pp$ interactions. 
The distant lobe shocks, separated from the compact ejecta by tens of parsecs, mark the downstream TeV-electron acceleration zone.
}
\label{fig:schematic_ss433}
\end{figure}

Taken together, SS~433, one of the prototypical microquasars, is revealed as a multi-zone 
particle accelerator by these results (as shown in Figure~\ref{fig:schematic_ss433}).
The compact baryonic ejecta provide the PeV-proton acceleration and injection channel identified by the radio-resolved magnetic rigidity. This does not exclude an additional acceleration channel at the parsec-scale recollimation shocks.
The transported protons subsequently radiate through $pp$ interactions in the neutral-gas-rich W50 environment, while the lower-energy TeV component resolved by H.E.S.S. is associated with downstream jet-shock regions where relativistic electrons are accelerated or reprocessed \citep{Abeysekara2018SS433,HESS2024SS433}.


\clearpage

\section{Methods}\label{Methods}

\subsection{VLBI data analysis}
\label{subsec:vlbi_magnetic_field}

The VLBI analysis of SS\,433 uses archival VLBA+Y1 data observed on 1998 June 16 (project code: BP042d) at 5, 8.4, and 15~GHz, providing simultaneous multi-frequency coverage of the jet to constrain the synchrotron spectrum while resolving its transverse width on milliarcsecond scales.
These data were calibrated in the Astronomical Image Processing System (AIPS; \citep{greisen2003aips}) and iteratively self-calibrated in phase and amplitude in \textsc{difmap} \citep{1997ASPC..125...77S}.
For spectral comparison across frequencies, all three visibility datasets were re-imaged and restored with a common circular Gaussian beam of $\Theta_{\rm maj}=3.537$\,mas (i.e., the major axis of the 5~GHz synthesized beam) and resampled onto an identical grid ($0.06$\,mas\,pixel$^{-1}$, $1024\times1024$ pixels).
Accurate spatial alignment of the multi-frequency maps is required to account for the core-shift effect.
Using the recent core-shift and jet-geometry measurements of SS\,433 by \citet{Yan2026SS433Collimation}, we aligned the 1998 images to the central binary system position with the exact offsets provided in their Table~2, which were derived from the alignment of optically thin jet components across frequencies.
The resulting VLBI images on 1998 June 16 are shown in Figure~\ref{fig:1998_3freq}.

The western (receding) jet of SS\,433 is significantly affected by free-free absorption (FFA; \citet{Paragi1999}), which strongly suppresses the low-frequency flux density, leading to inaccurate radio spectral-index measurements and hence unreliable equipartition magnetic-field estimates.
Therefore, we restrict the analysis to the eastern (approaching) jet.
For the eastern jet, the 8.4~GHz image is selected as the reference frequency for tracing the jet and measuring its transverse geometry, providing a good compromise between sensitivity and angular resolution.
Next, we traced the jet ridge line on the 8.4~GHz image outward from the core, starting at a position angle of $100^\circ$ in steps of $\Theta_{\rm maj}/4 = 0.88$\,mas.
At each step, the ridge point was set by fitting the transverse brightness profile with a Gaussian model \citep{Condon1997}:
\begin{equation}
I_\nu(u)=I_{0}\exp\left[-\frac{(u-u_{0})^2}{2\sigma^2}\right]+I_{\rm bg}.
\label{eq:transverse_profile}
\end{equation}
Here, $u$ is the transverse offset from the local jet axis, $I_{0}$ is the peak brightness of the Gaussian component above the local background, the peak position $u_0$ defines the ridge point, $\sigma$ is the transverse Gaussian standard deviation, and $I_{\rm bg}$ is a local constant background which absorbs residual large-scale emission and noise pedestal.
The transverse direction $u$ at each step is taken perpendicular to the local jet axis, which is updated iteratively from successive ridge points so that the slice geometry follows the jet curvature.
The tracing stopped where the transverse peak fell below $3\,I_{\rm rms}$.
At each ridge point, we integrated the flux within an aperture of $\pm\Theta_{\rm maj}/2$ along the jet and $\pm1.6\sigma$ across it (capturing ${\sim}89\%$ of the flux while rejecting neighbouring transverse structure), applying the identical aperture to the 5, 8.4 and 15\,GHz images to obtain matched flux densities.
The deconvolved width $W = (\Theta_{\rm fit}^{2} - \Theta_{\rm maj}^{2})^{1/2}$, with the fitted FWHM $\Theta_{\rm fit} = 2\sqrt{2\ln 2}\,\sigma$, was retained only where the jet was transversely resolved ($\Theta_{\rm fit} > \Theta_{\rm maj}$) with a fitted amplitude exceeding $20\,I_{\rm rms}$.
Since the jet is marginally resolved ($\Theta_{\rm fit}$ exceeds the beam by 5--15\%), a systematic uncertainty of $\Theta_{\rm maj}/10$ is assigned to $W$.
We adopted a viewing angle of $\theta_{\rm view} = 57^\circ$ for the 1998 epoch \citep{Yan2026SS433Collimation} to compute the deprojected distances, so that $H = R/\sin\theta_{\rm view}$.
At $D = 5.5$~kpc, 1~mas corresponds to a projected separation of 5.50~AU, or 6.56~AU deprojected.

Figure~\ref{fig:1998_spectral_index} shows the two-point power-law spectral indices $\alpha_{5}^{8.4}$ and $\alpha_{8.4}^{15}$ ($S\propto\nu^{-\alpha}$) for all 22 transverse slices along the eastern jet.
The three innermost slices ($H\lesssim\Theta_{\rm maj}$) display an inverted 5--8.4\,GHz spectrum ($\alpha_{5}^{8.4}<0$), consistent with the residual free-free absorption of the 5\,GHz emission in the immediate vicinity of the core.
In contrast, the outermost slices ($H\gtrsim 12$\,mas) carry large uncertainties because the 15\,GHz signal-to-noise ratio within the aperture is low, leaving the spectral indices essentially unconstrained.
In the intervening mid-jet, ten consecutive slices spanning $H\approx 3.5$--$12$\,mas show $\alpha_{5}^{8.4}$ and $\alpha_{8.4}^{15}$ to be in close agreement, which is consistent with the typical value of $\sim0.6$ for optically thin synchrotron emission.
For the magnetic-field measurements, we retain only those slices that satisfy the following criteria: 
\begin{itemize}
\item deprojected distance $H\gtrsim\Theta_{\rm maj}$, beyond which the transverse Gaussian fit is no longer beam-limited and the slice is free of core contamination;
\item a fitted Gaussian peak amplitude exceeding $20\,I_{\rm rms}$, and a $15$\,GHz integrated flux density above $5\sigma_{\rm rms}$ within the aperture;
\item optically thin spectral index $0.3<\alpha_{8.4}^{15}<1.0$;
\item ridge self-consistency, defined as the ratio of cumulative arc-length to straight-line distance from the core remaining below $1.5$, which rejects spurious points where the ridge loops back on itself.
\end{itemize}
These cuts yield 10 clean slices spanning $H=3.6$--$11.5$\,mas (${\approx}24$--$75$\,AU deprojected) as shown in Figure~\ref{fig:1998_slices}.

\subsection{Equipartition magnetic rigidity}

Based on the aperture-integrated flux densities and fitted angular sizes, the observed brightness temperatures across the eastern-jet slices are $T_{\rm b} \sim 10^{7}-10^{8}$\,K. 
This lies two to three orders of magnitude below the equipartition brightness temperature at the synchrotron self-absorption (SSA) turnover, $T_{\rm b,eq}\simeq5\times10^{10}$\,K \citep{1994ApJ...426...51R}. 
Such low $T_{\rm b}$ values are expected given the large viewing angle of the SS\,433 jet ($\theta_{\rm view}\simeq57^\circ$), which lacks significant Doppler boosting.
Consequently, the SSA turnover is not boosted into the observing frequencies (5--15\,GHz), unlike in the beamed, self-absorbed AGN cores, where the SSA magnetic field is routinely measured \citep{2024A&A...685L..11G,2026ApJ..1002...52X}.
In addition, no significant linear polarization is detected in the VLBA images, likely owing to strong depolarization near the core.
In this case, the classical equipartition (minimum-energy) argument provides our per-slice estimate of the magnetic field along the jet.

For optically thin synchrotron-emitting components, the equipartition magnetic field strength is given by \citep{Pacholczyk1970, BeckKrause2005}:
\begin{equation}
B_{{\rm eq}}=\left[\frac{6\pi(1+k)c_{12}L_{{\rm syn}}}{\phi V}\right]^{2/7}.
\label{eq:beq_classical}
\end{equation}
Here, $k$ is the ratio of non-radiating to radiating particle energy.
We adopt $k=1$ as the reference value and assess the sensitivity of $B_{\rm eq}$ to the poorly constrained hadronic energy content by also considering cases $k=0$ and $k=100$.
The volume filling factor $\phi$ is set to unity, following the standard equipartition convention \citep{Pacholczyk1970}.
The constant $c_{12}$ depends on the spectral index $\alpha$ and the integration frequency range $(\nu_1,\nu_2)$:
\begin{equation}
c_{12} = c_1^{1/2}\,c_2^{-1}\left[\frac{2\alpha-2}{2\alpha-1}\right]
\frac{\nu_1^{(1-2\alpha)/2}-\nu_2^{(1-2\alpha)/2}}{\nu_1^{\,1-\alpha}-\nu_2^{\,1-\alpha}},
\end{equation}
where $c_1=6.27\times10^{18}$ and $c_2=2.37\times10^{-3}$ are the synchrotron constants tabulated in \citet{Pacholczyk1970} (cgs units).
The analytic expression reproduces the Pacholczyk tabulation within ${\sim}5$--$10\%$.
It is indeterminate at $\alpha=0.5$, where we interpolate linearly between the analytic values at $\alpha=0.44$ and $0.56$.
We adopt $(\nu_1,\nu_2) = (10\,{\rm MHz}, 100\,{\rm GHz})$ as the integration limits.
Synchrotron luminosity $L_{\rm syn}$ follows from integrating the power-law spectrum $S_\nu\propto\nu^{-\alpha}$ over the same frequency range:
\begin{equation}
L_{\rm syn} = 4\pi D^{2}\,S_0\,\nu_0^{\alpha}\,\frac{\nu_u^{\,1-\alpha}-\nu_l^{\,1-\alpha}}{1-\alpha},
\label{eq:lsyn}
\end{equation}
where $S_0$ is the flux density at $\nu_0=8.4$\,GHz and $D=5.5$\,kpc is the distance to SS\,433.
Each slice is modelled as a cylinder with transverse radius $W/2$ and deprojected length $\Theta_{\rm maj}/\sin\theta_{\rm view}$ (one beam along the jet):
\begin{equation}
V = \pi\left(\frac{W}{2}\right)^{2}\frac{\Theta_{\rm maj}}{\sin\theta_{\rm view}}.
\label{eq:V}
\end{equation}

The equipartition magnetic field strength of each slice, $B_{\rm eq}$, is computed from the locally measured synchrotron luminosity $L_{\rm syn}$, deconvolved width $W$, and optically thin spectral index $\alpha_{8.4}^{15}$.
Uncertainties of each slice are propagated by Monte Carlo, sampling the flux density (thermal noise plus a $10\%$ amplitude-calibration term), the width ($\pm\Theta_{\rm maj}/10$), and the spectral index (thermal-noise-dominated, as the common calibration largely cancels in the flux ratio).
The quoted per-slice errors are the $68\%$ ($1\sigma$) intervals of the resulting $B_{\rm eq}$ distributions.
A least-squares power-law fit to the ten slices yields $B_{\rm eq}\propto H^{-0.50\pm0.12}$, where the quoted uncertainty is derived from a Monte Carlo analysis that treats the amplitude calibration and the width systematic as coherent across all slices.
Over the ten slices, the deconvolved width expands as $W\propto H^{0.44\pm0.12}$.
The equipartition field then scales with the jet width as $B\propto W^{-1.1\pm0.3}$.
Magnetic flux conservation in a jet of width $W$ at roughly constant speed predicts $B_{\rm t}\propto W^{-1}$ for a toroidal field and $B_{\rm p}\propto W^{-2}$ for a poloidal field \citep{1984RvMP...56..255B,LiangChen2021ApJ906105}. The measurement aligns more closely with the toroidal scaling and disfavours the poloidal one at $\sim3\sigma$. 
This is consistent with the millimetre polarimetry on considerably larger scales, which finds field vectors perpendicular to the jet ridge line within ${\sim}0.35$\,arcsec of the core, as expected for a toroidal configuration \citep{2018ApJ...867L..25B,lopez2026magnetic}.
On further out scales, the magnetic field in SS\,433 transitions to a poloidal configuration (parallel to the ridge line), consistent with the region where colliding ejecta are expected to reorder the field \citep{Roberts2008SS433,2018ApJ...867L..25B}. 
The jet trace of SS\,433 follows a helical path characterized by a $\sim162$-day precession period and a precession cone half-opening angle of $20^\circ\!.92 \pm 0^\circ\!.08$ \citep{2001ApJ...561.1027E}.
Our milliarcsecond-scale VLBI slices are taken perpendicular to the local jet ridge, which remains approximately straight over the sampled range, so the measured profiles are free from contamination by the projected helical geometry.
We define the local magnetic rigidity scale as $B(H)\,R_{\rm acc}(H)$, where $R_{\rm acc}$ is the transverse size of the acceleration region. 
In relativistic jets, the spatial scale of the particle acceleration region need not strictly trace the apparent synchrotron emitting profile \citep{2023MNRAS.525.5298H,2026A&A...708A.365W}. We therefore adopt a first-order conical scaling $R_{\rm acc}=\alpha_{\rm j}H$ as a fiducial baseline \citep{Hillas1984,RiegerBoschRamonDuffy2007,KoteraOlinto2011}, which we treat as a phenomenological parameterization rather than a direct measurement of the jet emitting geometry.

\subsection{Gamma-ray data analysis}
\label{subsec:gamma_sed_fitting}

We fitted the LHAASO spectral measurements~\citep{LHAASO2025Microquasars}
with \textsc{Naima}~\citep{Zabalza2015Naima}. Following the LHAASO
interpretation that the emission below 30\,TeV may be predominantly leptonic,
whereas a hadronic contribution is required at higher energies, we considered
two fitting ranges, $>1\,\mathrm{TeV}$ and $>30\,\mathrm{TeV}$, and tested
both pure hadronic and leptohadronic models. The latter allows for a residual
leptonic contribution when constraining the maximum proton energy,
$E_{p,\max}$.

The parent particle distributions are parametrized as
\begin{equation}
\frac{dN_i}{dE_i}
=
A_i
\left(\frac{E_i}{E_{0,i}}\right)^{-\alpha_i}
\exp\left[
-\left(\frac{E_i}{E_{i,\max}}\right)^{\beta_i}
\right],
\qquad i=p,e,
\label{eq:particle_ecpl}
\end{equation}
with $\beta_p=1$, and $\beta_e=2$.
In the pure hadronic model, the gamma ray emission is produced by neutral
pion decay following $pp$ interactions. The target density rescales the
required proton energy content but does not affect the fitted spectral
cutoff. In the leptohadronic model, the same hadronic component is
supplemented by inverse Compton emission on the cosmic microwave background.
We fix $E_{e,\max}=200\,\mathrm{TeV}$ following the H.E.S.S. measurements~
\citep{HESS2024SS433} and adopt a Gaussian prior
$\alpha_e=2.35\pm0.15$, consistent with the electron indices inferred by
LHAASO~\citep{LHAASO2025Microquasars} and H.E.S.S.~
\citep{HESS2024SS433}. Relative model preference is assessed using the
Bayesian information criterion (BIC)~\citep{schwarz1978estimating}.

For the pure hadronic model, the $>1\,\mathrm{TeV}$ fit gives
$\alpha_p=2.39^{+0.12}_{-0.15}$ and
$E_{p,\max}=1.64^{+2.16}_{-0.73}\,\mathrm{PeV}$
($\mathrm{BIC}=16.47$), while the $>30\,\mathrm{TeV}$ fit gives
$\alpha_p=2.74^{+0.29}_{-0.34}$ and
$E_{p,\max}=2.34^{+3.70}_{-1.34}\,\mathrm{PeV}$
($\mathrm{BIC}=9.76$).
For the leptohadronic model, the corresponding fits give
$\alpha_p=2.79^{+0.77}_{-0.32}$ and
$E_{p,\max}=3.08^{+4.04}_{-1.97}\,\mathrm{PeV}$
($\mathrm{BIC}=20.44$) above $1\,\mathrm{TeV}$, and
$\alpha_p=2.63^{+0.41}_{-0.39}$ and
$E_{p,\max}=2.63^{+4.00}_{-1.57}\,\mathrm{PeV}$
($\mathrm{BIC}=13.65$) above $30\,\mathrm{TeV}$.
We therefore adopt the lepto-hadronic model as the fiducial physical interpretation, because it accommodates distinct origins for the TeV and UHE emissions, while noting that the pure hadronic model yields a lower BIC. The relatively broad uncertainties, particularly for the lepto-hadronic fits, reflect degeneracy between the two spectral components and the limited statistical leverage of the current data. Nevertheless, the inferred proton cutoff remains in the PeV regime for both emission models and both fitting ranges, which is the quantity relevant to the magnetic-rigidity analysis.
For the energy budget calculations below, we also use the LHAASO spectrum above 100\,TeV,
\begin{equation}
\frac{dN}{dE}
=
(3.45\pm1.61)\times10^{-16}
\left(\frac{E}{50\,\mathrm{TeV}}\right)^{-3.96\pm0.25}
\,\mathrm{TeV}^{-1}\,\mathrm{cm}^{-2}\,\mathrm{s}^{-1},
\end{equation}
which corresponds to an energy flux
$F_{\gamma}(>100\,\mathrm{TeV})
\simeq1.81\times10^{-13}\,
\mathrm{erg\,cm^{-2}\,s^{-1}}$.

\subsection{Hillas confinement}
\label{subsec:acceleration_methods}

For each radio component, we test whether the local magnetic rigidity is
sufficient to confine protons at the energies required by the gamma ray
spectrum~\citep{Hillas1984,KoteraOlinto2011}. The distance $H_i$ is measured
along the deprojected radio ridge line. As a fiducial transverse scale of the
acceleration region, we adopt
\begin{equation}
R_{{\rm acc},i}\simeq \alpha_j H_i ,
\label{eq:racc_alphaH_methods}
\end{equation}
with $\alpha_j=0.1$. Here, $R_{\rm acc}$ represents the transverse extent
available for particle confinement and is not identified with the synchrotron
FWHM. The sensitivity to the observed jet width is assessed separately below.
The Hillas condition requires $r_{\rm L}<R_{\rm acc}$, with
$r_{\rm L}=E_p/(ZeB)$ for relativistic particles. At
$H=100\,{\rm AU}$, a $2.63\,{\rm PeV}$ proton in a
$0.19\,{\rm G}$ field has $r_{\rm L}\simeq3.1\,{\rm AU}$, below the
fiducial $R_{\rm acc}\simeq10\,{\rm AU}$. Accounting for the finite jet
speed, we define the relevant Hillas scale as
$E_{{\rm H},\beta}\simeq\beta_jZeBR_{\rm acc}$, adopting $Z=1$ and
$\beta_j\simeq0.26$.

The radio measurements give
\[
B(H)=B_{100}
\left(\frac{H}{100\,{\rm AU}}\right)^{-q},
\]
with $B_{100}\simeq0.19\,{\rm G}$ and $q=0.50\pm0.12$. Combining this
profile with $R_{\rm acc}=\alpha_jH$ gives
\begin{equation}
E_{{\rm H},\beta}(H)
\simeq
2.20
\left(\frac{B_{100}}{0.19\,{\rm G}}\right)
\left(\frac{\alpha_j}{0.1}\right)
\left(\frac{\beta_j}{0.26}\right)
\left(\frac{H}{100\,{\rm AU}}\right)^{1-q}
{\rm PeV}.
\label{eq:epmax_profile_methods}
\end{equation}
Since $q<1$, the product $B(H)H$ increases with distance and
$E_{{\rm H},\beta}\propto H^{0.50\pm0.12}$. The corresponding value is
$\simeq2.2\,{\rm PeV}$ at $100\,{\rm AU}$ and reaches the
$2.63\,{\rm PeV}$ proton cutoff inferred from the gamma ray fit at a
nominal distance of $H\simeq140\,{\rm AU}$.
The value of this crossing distance depends on the adopted transverse scale and should not be interpreted as an exact localization of the accelerator. The more robust result is that the measured field decreases substantially more slowly than $H^{-1}$, allowing the available magnetic rigidity to reach the PeV regime on scales of hundreds of AU. Radio imaging suggests that the SS~433 ejecta may undergo deceleration on arcsecond scales \citep{Stirling2004SS433,Panferov2014SS433}. Adopting the phenomenological prescription considered by \citet{Stirling2004SS433}, together with $B\propto H^{-0.50}$ and $R_{\rm acc}\propto H$, gives $E_{\rm H,\beta}\propto\beta(H)H^{1/2}$ and a nominal turnover at $H_{\rm crit}\sim0.1\,{\rm pc}$. This value should be regarded as a characteristic scale, while the exact boundary is not directly constrained by current observations.

We also tested the sensitivity of the Hillas estimate to the adopted transverse scale using the FWHM-based proxy $R_W=W/2$.
The ten slices give $W\propto H^{0.44\pm0.12}$, consistent with the collimation slope $m=0.53\pm0.08$ reported by \citet{Yan2026SS433Collimation}. Fixing $m=0.53$ and normalizing this relation to the ten measured widths gives $W_{100}
=13.1\,{\rm AU}$ and hence $R_W(100\,{\rm AU})=6.5\,{\rm AU}$. This is an extrapolation from the observed range; no
inclination correction is required for the transverse width. For $B=0.19\,{\rm G}$, this scale gives $E_{{\rm H},\beta}
=1.45\,{\rm PeV}$, while pure confinement remains satisfied because $r_{\rm L}=3.1\,{\rm AU}<R_W$. For the motivated
baryon-loading range $k=20$--60, the corresponding field of $0.37$--$0.50\,{\rm G}$ raises the benchmark to $E_{{\rm H},
\beta}=2.8$--$3.8\,{\rm PeV}$. 

The Poynting-flux bound derived above provides a complementary check. At fixed magnetization, one-sided jet power and speed, and for a field filling the adopted cross-section, $B\propto R_{\rm acc}^{-1}$. The limiting product $BR_{\rm acc}$ is therefore independent of the adopted transverse scale. Geometry changes the numerical speed-weighted energy and the nominal crossing distance, but both transverse prescriptions support a PeV-scale confinement condition. Satisfying the Hillas condition shows that the local ejecta have sufficient magnetic rigidity, but PeV-proton injection also requires acceleration before advection or diffusive escape removes the particles from the region.

\subsection{Particle acceleration timescale}
\label{subsec:acceleration_timescale}

The Hillas condition establishes whether the radio ejecta can confine protons
at the energies required by the gamma ray spectrum, but does not ensure that
these energies can be reached before particle removal. We therefore compare
the acceleration time with the characteristic flow and escape times.
For first order Fermi acceleration, we adopt
$t_{\rm acc,I}\simeq\eta_{\rm I}(r_{\rm L}/c)\beta_{\rm sh}^{-2}$,
where $\beta_{\rm sh}$ is the shock speed in units of $c$ and
$\eta_{\rm I}$ accounts for the diffusion and shock properties. We take
$\eta_{\rm I}=1$ as a Bohm limit and $20/3$ as a standard
parallel shock Bohm benchmark~\citep{Drury1983,BlandfordEichler1987,
RiegerBoschRamonDuffy2007}. For
$E_p=2.63\,{\rm PeV}$, $B=0.19\,{\rm G}$, and
$\beta_{\rm sh}\simeq\beta_j\simeq0.26$, this gives
\begin{equation}
t_{\rm acc,I}
\simeq
0.26\,\eta_{\rm I}
\left(\frac{E_p}{2.63\,{\rm PeV}}\right)
\left(\frac{B}{0.19\,{\rm G}}\right)^{-1}
\left(\frac{\beta_{\rm sh}}{0.26}\right)^{-2}
{\rm d}.
\label{eq:tacc_fermi1_numeric}
\end{equation}
The corresponding acceleration times are
$\simeq0.26\,{\rm d}$ for $\eta_{\rm I}=1$ and
$\simeq1.7\,{\rm d}$ for $\eta_{\rm I}=20/3$.
The characteristic flow time is
\begin{equation}
t_{\rm dyn}
\simeq
\frac{H}{\beta_j c}
\simeq
2.2
\left(\frac{H}{100\,{\rm AU}}\right)
\left(\frac{\beta_j}{0.26}\right)^{-1}
{\rm d},
\label{eq:tdyn_numeric}
\end{equation}
where we adopt $\beta_j\simeq0.26$ for the SS~433 jets~
\citep{Margon1984,Fabrika2004,Sakemi2021}. At
$H\simeq100\,{\rm AU}$, both acceleration benchmarks therefore satisfy
$t_{\rm acc,I}<t_{\rm dyn}$. The measured field profile further gives
$t_{\rm acc,I}\propto B^{-1}\propto H^{0.50\pm0.12}$, while
$t_{\rm dyn}\propto H$, yielding
$t_{\rm acc,I}/t_{\rm dyn}\propto H^{-0.50\pm0.12}$.
Acceleration relative to advection therefore becomes progressively less
restrictive downstream.

Particle escape provides the less certain constraint. We estimate
\begin{equation}
t_{\rm esc}
\simeq
0.28\,\chi^{-1}
\left(\frac{l_{\rm esc}}{10\,{\rm AU}}\right)^2
\left(\frac{E_p}{2.63\,{\rm PeV}}\right)^{-1}
\left(\frac{B}{0.19\,{\rm G}}\right)
{\rm d},
\label{eq:tesc_numeric}
\end{equation}
using $D(E)=\chi D_{\rm B}(E)=\chi r_{\rm L}c/3$, with $\chi\geq1$.
For transverse escape over $l_{\rm esc}=R_{\rm acc}\simeq10\,{\rm AU}$,
even the Bohm case gives $t_{\rm esc}\simeq0.28\,{\rm d}$, comparable to
the fastest first order acceleration time. Larger $\chi$ would shorten the
escape time further, making this the most restrictive transport geometry.
If particles instead remain confined over $l_{\rm esc}\sim H$, the residence
time increases by approximately $(H/R_{\rm acc})^2$, substantially relaxing
the constraint. The dominant uncertainty is therefore the transport geometry
and degree of particle confinement. Stochastic second order Fermi
acceleration may provide an additional channel in strongly turbulent ejecta,
but should not required for the fiducial scenario~
\citep{BlandfordEichler1987,RiegerBoschRamonDuffy2007}.
Combining the confinement and timescale constraints, the compact ejecta can
serve as PeV proton injection sites where
$E_{{\rm H},\beta}\gtrsim E_{\rm cut}^{p}$ and the acceleration time remains
shorter than the relevant residence time. The exact location depends on the
adopted transverse scale, while the increasing magnetic rigidity follows from
the measured shallow field profile.

The same field profile strongly constrains a compact PeV electron origin for
the highest energy gamma rays. Extrapolating the measured profile to
$H\sim10^{3}\,{\rm AU}$ gives $B\simeq0.06\,{\rm G}$, for which the
synchrotron cooling time is
\begin{equation}
t_{\rm syn,e}
=
\frac{6\pi m_e^2c^3}{\sigma_{\rm T}B^2E_e}
\simeq
1.3\times10^{-3}
\left(\frac{E_e}{1\,{\rm PeV}}\right)^{-1}
\left(\frac{B}{0.06\,{\rm G}}\right)^{-2}
{\rm d}.
\label{eq:tsyn_e_numeric}
\end{equation}
A PeV electron therefore cools in only $\sim110\,{\rm s}$, far shorter than
the local flow time~\citep{RybickiLightman1979}. Moreover,
$U_B\simeq1.4\times10^{-4}\,{\rm erg\,cm^{-3}}$ exceeds the CMB energy
density by a factor of $\sim3.4\times10^8$, while inverse Compton scattering
of PeV electrons is already in the Klein Nishina regime for CMB photons and
is even more suppressed for higher energy target photons~
\citep{BlumenthalGould1970}.

These cooling constraints disfavor the compact radio ejecta as the dominant
PeV electron inverse Compton radiation zone for the $>100\,{\rm TeV}$
emission. They do not exclude lower energy electrons elsewhere in SS~433,
including those associated with the downstream parsec scale jet shocks~
\citep{HESS2024SS433}. The compact ejecta are therefore more naturally
identified with PeV proton acceleration and injection, while the highest
energy radiation is produced farther downstream or in the surrounding
environment.

Photohadronic losses are also negligible on the radio-ejecta scales considered here. The relevant channel is $p+\gamma \rightarrow \Delta^{+}$, followed by either $\Delta^{+}\rightarrow p+\pi^{0}$ with $\pi^{0}\rightarrow 2\gamma$, or $\Delta^{+}\rightarrow n+\pi^{+}$ with $\pi^{+}\rightarrow \mu^{+}+\nu_{\mu}\rightarrow e^{+}+\nu_{e}+\bar{\nu}_{\mu}+\nu_{\mu}$ \citep{kelner2008energy}. A $1\,{\rm keV}$ target photon field lowers the photomeson threshold to $E_{p,\rm thr}\sim0.1$--$0.3\,{\rm PeV}$, so the PeV protons considered here are well above threshold. However, the loss rate is controlled by the photon number density rather than by the threshold alone. Even for an isotropic $L_{\rm ph}=10^{39}\,{\rm erg\,s^{-1}}$ radiation field with characteristic photon energy $\epsilon_{\gamma}=1\,{\rm keV}$, geometric dilution at $H=100\,{\rm AU}$ gives $n_{\gamma}\simeq7\times10^{5}\,{\rm cm^{-3}}$ and $t_{p\gamma}\sim10^{4}\,{\rm yr}$. This exceeds both the near-Bohm acceleration time and the local flow time by more than six orders of magnitude. Thus $p\gamma$ interactions may be relevant only in a sub-AU corona or disk-funnel region where an intense X-ray field is co-spatial with the accelerator; they do not affect the PeV-proton injection condition inferred from the $\sim100\,{\rm AU}$ radio ejecta.

\subsection{Energy budget}
\label{subsec:energy_budget}

The resolved radio emission constrains the magnetic field in the compact ejecta, while the jet power provides an independent test of its energetic plausibility. We define
$k\equiv U_{\rm nonrad}/U_e$ as the ratio of nonradiating to synchrotron emitting particle energy densities. Using the observed synchrotron luminosity and emitting volume, the classical minimum energy estimate at $H=100\,{\rm AU}$ is
$B_{\rm eq}(k)
\simeq
0.19
\left(\frac{1+k}{2}\right)^{2/7}
{\rm G}$,
following Equation~\ref{eq:beq_classical}~\citep{BeckKrause2005}. This provides an observational reference for the minimum total energy state, but is not a strict lower bound on the field, since both particle dominated and magnetically dominated configurations can depart from equipartition.

Under the minimum energy condition, $U_{\rm part}\simeq4U_B/3$, and the corresponding internal energy flux through a region of radius $R_{\rm acc}$ is
$L_{\rm int,loc}
\simeq
\frac{7}{24}R_{\rm acc}^{2}\beta_j c B_{\rm eq}^{2}$.
For $R_{\rm acc}=10\,{\rm AU}$ and $\beta_j=0.26$, the range $k=20$ to $60$ gives $B_{\rm eq}=0.37$ to $0.50\,{\rm G}$. This corresponds to only $0.7\%$ to $1.3\%$ of a total system jet power
$L_{\rm k,sys}=10^{39}\,{\rm erg\,s^{-1}}$, or $2.4\%$ to $4.3\%$ of the power of one jet,
$L_{\rm k,1}\simeq3\times10^{38}\,{\rm erg\,s^{-1}}$.
The inferred subgauss fields are therefore readily accommodated by the available jet power. Increasing $k$ raises the field only weakly,
$B_{\rm eq}\propto(1+k)^{2/7}$, while the required internal energy flux scales as $(1+k)^{4/7}$.

The value of $k$ is not directly measured. Estimates based on shock accelerated particle populations give proton to electron number ratios
$K_0\simeq40$ to $90$ for synchrotron indices
$\alpha_0\simeq0.5$ to $0.6$~\citep{BeckKrause2005}, corresponding to baryonic loading of order $k\sim20$ to $60$ for plausible particle distributions and low energy cutoffs. An independent consistency check is provided by the core shift estimate
$B_{\rm cs}\simeq0.4\,{\rm G}$ at $35\,{\rm AU}$~\citep{Paragi1999}, close to the $k=1$ extrapolation of the resolved profile,
$B_{\rm eq}\simeq0.32\,{\rm G}$. Because the measurements correspond to different epochs and emitting regions, this comparison constrains the plausible field scale but does not determine $k$.

A constraint independent of minimum energy follows from the Poynting flux. For a predominantly transverse field filling the adopted acceleration region,
$L_B=\frac{B^2}{4\pi}\beta_j c\pi R_{\rm acc}^2
\equiv
\sigma L_{\rm k,1}$,
which gives
\begin{equation}
B_{\sigma}
=
0.83\,{\rm G}
\left(\frac{\sigma}{0.1}\right)^{1/2}
\left(\frac{L_{\rm k,1}}
{3\times10^{38}\,{\rm erg\,s^{-1}}}\right)^{1/2}
\left(\frac{R_{\rm acc}}{10\,{\rm AU}}\right)^{-1}
\left(\frac{\beta_j}{0.26}\right)^{-1/2}.
\label{eq:b_sigma}
\end{equation}
Here $\sigma=L_B/L_{\rm k,1}$ is the fraction of the power of one jet carried as Poynting flux. We adopt
$L_{\rm k,1}\simeq3\times10^{38}\,{\rm erg\,s^{-1}}$~
\citep{fogantini2023nustar,marshall2002high}; relativistic corrections are small, with $\Gamma_j^2\simeq1.07$ at $\beta_j=0.26$. A magnetic fraction $\sigma=0.1$ permits
$B\simeq0.83\,{\rm G}$ across a $10\,{\rm AU}$ region, while $\sigma=1$ gives an upper envelope of approximately $2.62\,{\rm G}$.

Combining the Poynting flux constraint with the Hillas condition removes the explicit dependence on $R_{\rm acc}$:
\begin{equation}
E_{\rm H,\beta}^{(\sigma)}
=
30.6\,Z
\left(\frac{\sigma}{1.0}\right)^{1/2}
\left(\frac{L_{\rm k,1}}
{3\times10^{38}\,{\rm erg\,s^{-1}}}\right)^{1/2}
\left(\frac{\beta_j}{0.26}\right)^{1/2}
{\rm PeV}.
\label{eq:hillas_sigma}
\end{equation}
Reaching $2.63\,{\rm PeV}$ therefore requires only
$\sigma\gtrsim7.4\times10^{-3}$ for the adopted jet power. This provides an independent check that the PeV magnetic rigidity inferred from the radio profile does not require an energetically extreme field configuration. Since $B\propto R_{\rm acc}^{-1}$ at fixed Poynting flux, the limiting product $BR_{\rm acc}$ is independent of the adopted transverse scale. The radio measurement and jet power constraint therefore provide complementary bounds on magnetic configurations capable of supporting PeV proton acceleration.

\subsection{Radiative interpretation}
\label{subsec:radiative_template_methods}

The acceleration region need not coincide with the gamma ray radiation
volume. The LHAASO emission above $100\,\mathrm{TeV}$ is concentrated toward
the central SS~433 region and overlaps with the H\,{\sc i} distribution~
\citep{LHAASO2025Microquasars}. This morphology motivates a scenario in which
protons accelerated in the compact baryonic ejecta subsequently interact with
neutral gas in W50, producing the highest energy gamma rays through $pp$
interactions~\citep{Kelner2006,Kafexhiu2014}.

The characteristic proton cooling time is
\begin{equation}
t_{pp}
\simeq
5\times10^{7}
\left(\frac{n_{\rm H}}{1\,{\rm cm}^{-3}}\right)^{-1}
{\rm yr},
\end{equation}
following \citet{Kelner2006}. The LHAASO spectrum adopted in
Section~\ref{subsec:gamma_sed_fitting} gives
$F_E(>100\,\mathrm{TeV})\simeq1.81\times10^{-13}\,
{\rm erg\,cm^{-2}\,s^{-1}}$, corresponding to
$L_{\gamma}(>100\,\mathrm{TeV})\simeq6.6\times10^{32}\,
{\rm erg\,s^{-1}}$ at $d=5.5\,{\rm kpc}$.

We use the compact jet density only to test whether the radio ejecta themselves
could account for this luminosity. For a conical baryonic jet with
$R_j=\alpha_jH$, the kinetic power implies
\begin{equation}
n_j(H)
\simeq
3.6\times10^4
\left(\frac{L_j}{10^{39}\,{\rm erg\,s^{-1}}}\right)
\left(\frac{\alpha_j}{0.1}\right)^{-2}
\left(\frac{\beta_j}{0.26}\right)^{-3}
\left(\frac{H}{100\,{\rm AU}}\right)^{-2}
{\rm cm}^{-3},
\end{equation}
following \citet{Fabrika2004}. At $H\simeq100\,{\rm AU}$,
$t_{pp}\sim1.4\times10^3\,{\rm yr}$, far longer than the local flow time.
For an advection limited residence time,
$t_{\rm res}\sim t_{\rm dyn}\simeq2.2\,{\rm d}$, the interaction efficiency is
only $f_{pp,\rm comp}\sim4\times10^{-6}$.

For a proton power $L_p=\xi_{\rm CR}L_j$, the compact emission is bounded by
\begin{equation}
\begin{aligned}
F_{\gamma,\rm comp}^{pp}
&\lesssim
2\times10^{-14}
\left(\frac{f_\gamma}{0.17}\right)
\left(\frac{\xi_{\rm CR}}{0.1}\right)
\left(\frac{L_j}{10^{39}\,{\rm erg\,s^{-1}}}\right)
\left(\frac{n_{\rm H}}{3.6\times10^4\,{\rm cm}^{-3}}\right)
\\
&\quad\times
\left(\frac{t_{\rm res}}{2.2\,{\rm d}}\right)
\left(\frac{d}{5.5\,{\rm kpc}}\right)^{-2}
{\rm erg\,cm^{-2}\,s^{-1}} .
\end{aligned}
\end{equation}
For the fiducial parameters, this is approximately $10\%$ of the observed
energy flux above $100\,\mathrm{TeV}$. The estimate scales linearly with the
adopted density and residence time and is therefore a consistency test rather
than a precise luminosity prediction. Longer particle confinement would
increase the local interaction efficiency.

For advection dominated transport, $n_j\propto H^{-2}$ and
$t_{\rm dyn}\propto H$, giving $f_{pp,\rm comp}\propto H^{-1}$. Thus, the
increase in magnetic rigidity downstream does not imply a corresponding
increase in the local gamma ray efficiency. The compact ejecta can therefore
act as an acceleration and injection region while remaining inefficient as a
hadronic radiation zone. Protons escaping or advecting into the surrounding
H\,{\sc i} environment encounter a larger target mass and interaction volume,
providing a natural origin for the extended emission above
$100\,\mathrm{TeV}$. This separation between acceleration and radiation is
consistent with the observed central H\,{\sc i} association.

\subsection{Morphological transport diagnosis}
\label{subsec:morphological_simulation_test}

We use a projected transport calculation to test whether PeV protons injected
from the compact ejecta can reproduce the broad morphology of the LHAASO
$>100\,\mathrm{TeV}$ emission after propagation into the central W50
environment. The calculation is intended as a morphology diagnostic.

The neutral target distribution is derived from the HI4PI data
cube~\citep{HI4PI2016} using the standard optically thin H\,{\sc i}
conversion~\citep{KalberlaKerp2009}. We consider velocity intervals of
$33$ to $56$, $70$ to $90$, $72$ to $82$, and $20$ to
$100\,\mathrm{km\,s^{-1}}$, adopting the broadest interval for the reference
calculation. The H\,{\sc i} column density is converted to an approximate
target density using $n_{\rm H}=N_{\rm HI}/L_{\rm los}$ with
$L_{\rm los}=100\,\mathrm{pc}$. No additional smoothing is applied to the
gas map. The resulting template therefore represents a projected target
density and not a reconstruction of the full spatial gas distribution.

The injected proton population is described by a cutoff power law with
$s=2.2$ and $E_{\rm cut}=2.63\,\mathrm{PeV}$. Propagation is described by
$D_{\perp}(E)=D_0(E/E_0)^{\delta}$ and
$D_{\parallel}=\eta D_{\perp}$, with the parallel direction aligned with the
jet position angle $\mathrm{PA}=100^\circ$. We adopt
$D_0=1.5\times10^{28}\,\mathrm{cm^2\,s^{-1}}$ at
$E_0=1\,\mathrm{PeV}$, $\delta=1/3$, $t=10^4\,\mathrm{yr}$,
$v_{\rm adv}=0$, and $\eta=1$, giving
$\ell_{\rm diff}(1\,\mathrm{PeV})=44.6\,\mathrm{pc}$. Here $t$ represents
an effective propagation or residence time of the PeV proton population.

The proton normalization,
$W_p=1.29\times10^{49}\,\mathrm{erg}$, is chosen so that the
$20$ to $100\,\mathrm{km\,s^{-1}}$ template reproduces the observed
integral flux above $100\,\mathrm{TeV}$ for the adopted $L_{\rm los}$.
It is therefore a flux normalization for the morphology test and not an
independent measurement of the proton energy content. The propagated proton
distribution is weighted by the H\,{\sc i} target density and convolved with
a Gaussian point spread function with $\sigma_{\rm PSF}=0.2^\circ$.

For isotropic diffusion, the resulting template has
$R_{68}\simeq0.64^\circ$, an axis ratio of $1.02$, and a centroid
displacement of $0.055^\circ$ from SS~433. Alternative H\,{\sc i}
velocity intervals mainly affect the absolute normalization, reflecting the
dependence of the inferred proton content on the adopted gas template and
line of sight depth.

We then vary the diffusion anisotropy while keeping the gas map and transport
scale fixed. Increasing $\eta$ from 1 to 3 elongates the emission along the
jet direction (Extended Data Fig.~\ref{fig:eta_transport_triptych}), with the
axis ratio increasing from $\simeq1.02$ to $\simeq1.5$. Comparison with the
released LHAASO $>100\,\mathrm{TeV}$ significance map within a $3^\circ$
region gives a minimum near $\eta\simeq1.5$ under the adopted assumptions~
\citep{LHAASO2025Microquasars}. Because the comparison uses a correlated
significance map rather than a likelihood based on counts, exposure, and
background, we do not interpret this value as a measurement of the diffusion
tensor. Strongly anisotropic transport along the jet is disfavoured, whereas
approximately isotropic or mildly anisotropic propagation into the central
H\,{\sc i} reservoir remains compatible with the observed morphology.

\section*{Data availability}

The VLBA+Y1 observations analysed in this study are publicly available through the NRAO Science Data Archive (\url{https://data.nrao.edu/}) under project code BP042d. The H.E.S.S. significance maps, flux maps and spectral measurements of SS~433 are available from the H.E.S.S. auxiliary-data repository (\url{https://www.mpi-hd.mpg.de/HESS/pages/publications/auxiliary/2023_SS433/index.html}). The LHAASO microquasar data products are available from the LHAASO public-data repository (\url{https://english.ihep.cas.cn/lhaaso/pdl/202110/t20211026_286779.html}). The HI4PI spectral-line data and column-density maps are available through the CDS/VizieR archive (\url{https://cdsarc.cds.unistra.fr/viz-bin/cat/J/A+A/594/A116}).

\section*{Code availability}
The custom analysis code used to generate the principal results will be made available to editors and referees upon request. An archived version will be deposited in a public repository upon publication.

\section*{Acknowledgements}
This work was supported by the Tianshan Talent Training Program (Grant No. 2023TSYCCX0099) and the National Key R\&D Program of China (Grant Nos. 2024YFA1611503, 2022SKA0120102, and 2021YFA0718500). J.L., P.-H.T.T. and L.Y. were supported by the
National Natural Science Foundation of China (NSFC) under
grant No. 12273122, National Astronomical Data Center, the
Greater Bay Area, under grant No. 2024B1212080003, and
science research grant from the China Manned Space Project
under CMS-CSST-2025-A13. X.Y. was supported by the China Postdoctoral Science Foundation under grant Nos. 2025M773200 and 2026T190850, and by the Xinjiang Tianchi Talent Program. R.-Y.L. was supported by the NSFC (Grant No. 12593852) and the Basic Research Program of Jiangsu (Grant No. BK20250059). Y.-F.H. was supported by the NSFC (Grant No. 12233002) and the Xinjiang Tianchi Talent Program. W.X. was supported by China Scholarship Council (Grant No. 202504910180). This work was also partly supported by the Urumqi Nanshan Astronomy and Deep Space Exploration Observation and Research Station of Xinjiang (XJYWZ2303) and the Central Guidance for Local Science and Technology Development Fund (Grant No. ZYYD2026JD01). 

The Very Long Baseline Array is operated by the National Radio Astronomy Observatory. The National Radio Astronomy Observatory and Green Bank Observatory are facilities of the U.S. National Science Foundation operated under cooperative agreement by Associated Universities, Inc. Scientific results from data presented in this publication are derived from the VLBA project BP042d.
This work also made use of publicly available high-level data products released by the LHAASO and H.E.S.S. collaborations.

\section*{Author contributions}

J.L. and L.C. (Lang Cui) conceived the study and designed the research plan. J.L. and W.X. performed the main data analysis and physical calculations, interpreted the results, prepared the figures, and wrote the initial manuscript. L.C. (Lang Cui) and P.-H.T.T. supervised the research and provided important guidance on the scientific interpretation and presentation. L.Y. performed the gamma-ray spectral fitting and contributed to writing the manuscript. F.A., X.Y., R.-Y.L., L.C. (Liang Chen), N.C., L.J., A.T. and Y.-F.H. contributed to scientific discussions and provided comments that improved the analysis and presentation.  All authors reviewed and commented on the manuscript.

\section*{Competing interests}
The authors declare no competing interests.



\clearpage
\setcounter{table}{0}
\renewcommand{\tablename}{Extended Data Table}

\newgeometry{margin=1cm}
\begin{landscape}
\begin{table}[p]
\centering
\caption{Transverse slice measurements and equipartition magnetic field estimates for the eastern jet of SS~433 on 1998 June 16 (VLBA+Y1). 
Column (1): projected core distance $R$; 
Column (2): deprojected distance $H$; 
Column (3): position angle PA (from north through east); 
Column (4): fitted transverse FWHM $\Theta_{\rm fit}$; 
Column (5): beam-deconvolved width $W$ (uncertainty $\Theta_{\rm maj}/10$); 
Columns (6)--(8): flux densities at 5, 8.4, and 15\,GHz, with errors combining thermal noise and a 10\% amplitude calibration uncertainty in quadrature; 
Columns (9)--(10): two-point spectral indices with thermal-noise errors; 
Column (11): Pacholczyk constant; 
Column (12): synchrotron luminosity integrated over 10\,MHz--100\,GHz from the flux density at 8.4\,GHz; 
Column (13): emitting volume; 
Column (14): equipartition field strength $B_{\rm eq}$, with $1\sigma$ uncertainties from Monte Carlo sampling for $k=1$.}
\label{tab:slices}
\setlength{\tabcolsep}{3.5pt}
\renewcommand{\arraystretch}{1.3}
\begin{tabular}{cccccccccccccc}
\toprule
$R$ & $H$ & PA & $\Theta_{\rm fit}$ & $W$ & $S_{5}$ & $S_{8.4}$ & $S_{15}$ & $\alpha_{5}^{8.4}$ & $\alpha_{8.4}^{15}$ & $c_{12}$ & $L_{\rm syn}$ & $V$ & $B_{\rm eq}$ \\
(mas) & (AU) & (deg) & (mas) & (mas) & (mJy) & (mJy) & (mJy) &  &  & ($10^{7}$) & ($10^{31}$~erg~s$^{-1}$) & ($10^{42}$~cm$^3$) & (mG) \\
(1) & (2) & (3) & (4) & (5) & (6) & (7) & (8) & (9) & (10) & (11) & (12) & (13) & (14) \\
\midrule
 3.56 & 23.4 & 109.9 & $3.72\pm0.02$ & $1.14\pm0.35$ & $67.3\pm6.8$ & $58.9\pm5.9$ & $47.8\pm4.8$ & $+0.25\pm0.03$ & $+0.35\pm0.03$ & $1.08\pm0.08$ & $13.8\pm1.4$ & $2.39\pm1.55$ & $345^{+88}_{-51}$ \\
 4.45 & 29.2 & 109.5 & $3.75\pm0.03$ & $1.24\pm0.35$ & $61.8\pm6.2$ & $47.7\pm4.8$ & $35.6\pm3.7$ & $+0.50\pm0.03$ & $+0.48\pm0.04$ & $1.52\pm0.18$ & $10.0\pm1.0$ & $2.84\pm1.68$ & $326^{+72}_{-46}$ \\
 5.33 & 35.0 & 108.5 & $3.80\pm0.02$ & $1.38\pm0.35$ & $55.2\pm5.6$ & $39.0\pm4.0$ & $27.3\pm2.8$ & $+0.67\pm0.04$ & $+0.59\pm0.06$ & $2.01\pm0.35$ & $7.8\pm0.8$ & $3.52\pm1.87$ & $313^{+61}_{-42}$ \\
 6.21 & 40.7 & 107.5 & $3.86\pm0.01$ & $1.55\pm0.35$ & $49.4\pm5.0$ & $33.1\pm3.4$ & $22.0\pm2.3$ & $+0.77\pm0.05$ & $+0.68\pm0.07$ & $2.63\pm0.61$ & $6.6\pm0.7$ & $4.41\pm2.02$ & $303^{+55}_{-38}$ \\
 7.09 & 46.5 & 106.4 & $3.93\pm0.01$ & $1.71\pm0.35$ & $43.5\pm4.4$ & $27.9\pm2.9$ & $18.0\pm2.0$ & $+0.85\pm0.05$ & $+0.73\pm0.08$ & $3.13\pm0.89$ & $5.6\pm0.7$ & $5.41\pm2.25$ & $287^{+51}_{-38}$ \\
 7.96 & 52.2 & 105.2 & $3.95\pm0.01$ & $1.75\pm0.35$ & $35.5\pm3.6$ & $21.8\pm2.3$ & $14.1\pm1.6$ & $+0.94\pm0.07$ & $+0.72\pm0.11$ & $3.04\pm1.08$ & $4.4\pm0.6$ & $5.65\pm2.36$ & $262^{+49}_{-38}$ \\
 8.83 & 57.9 & 104.3 & $3.98\pm0.01$ & $1.82\pm0.35$ & $28.6\pm2.9$ & $17.1\pm1.8$ & $11.3\pm1.4$ & $+0.98\pm0.09$ & $+0.69\pm0.13$ & $2.79\pm1.20$ & $3.4\pm0.6$ & $6.09\pm2.44$ & $232^{+51}_{-35}$ \\
 9.73 & 63.8 & 104.4 & $4.03\pm0.03$ & $1.94\pm0.35$ & $23.6\pm2.4$ & $13.8\pm1.5$ &  $9.3\pm1.2$ & $+1.03\pm0.11$ & $+0.65\pm0.16$ & $2.42\pm1.29$ & $2.7\pm0.6$ & $6.94\pm2.56$ & $203^{+47}_{-31}$ \\
10.61 & 69.6 & 105.3 & $3.96\pm0.02$ & $1.79\pm0.35$ & $17.9\pm1.9$ & $11.2\pm1.3$ &  $7.4\pm1.1$ & $+0.90\pm0.13$ & $+0.70\pm0.20$ & $2.82\pm1.55$ & $2.2\pm0.9$ & $5.90\pm2.33$ & $207^{+53}_{-37}$ \\
11.51 & 75.5 & 105.3 & $3.95\pm0.02$ & $1.76\pm0.35$ & $14.8\pm1.6$ &  $9.7\pm1.2$ &  $5.8\pm1.0$ & $+0.81\pm0.16$ & $+0.86\pm0.25$ & $4.70\pm1.71$ & $2.2\pm2.1$ & $5.75\pm2.37$ & $216^{+51}_{-43}$ \\
\bottomrule
\end{tabular}
\end{table}
\end{landscape}
\restoregeometry

\clearpage
\setcounter{figure}{0}
\renewcommand{\figurename}{Extended Data Figure}

\begin{figure}[h]
\centering
\includegraphics[width=1\textwidth]{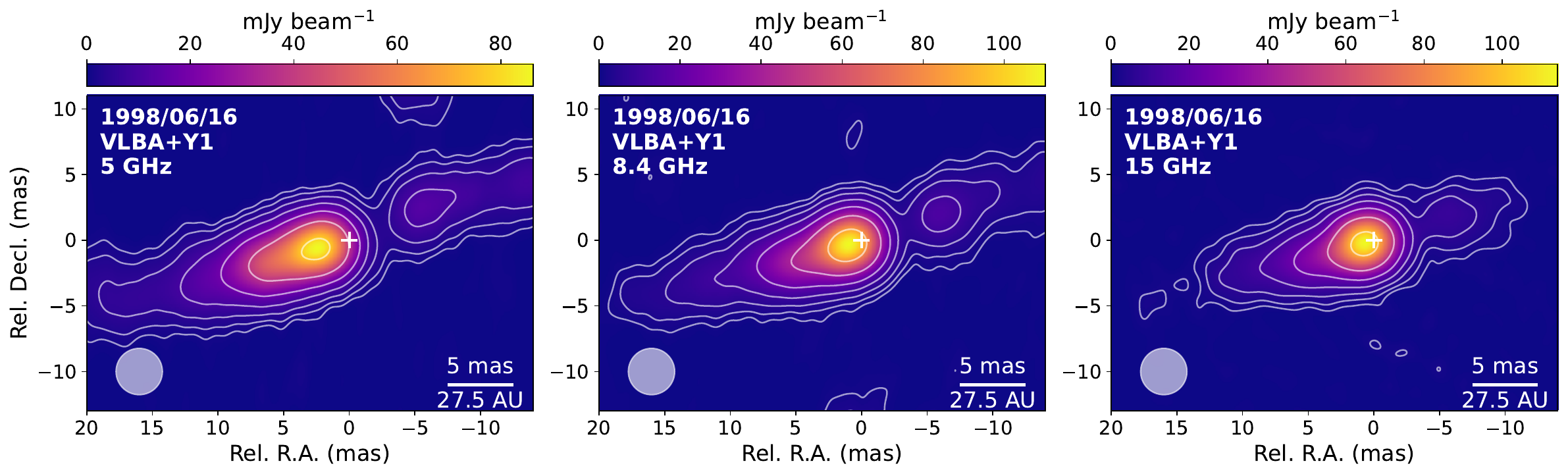}
\caption{Naturally weighted multi-frequency VLBI images of SS~433 on 1998 June 16. The background color scale represents the total intensity. All three maps are restored with a common circular Gaussian synthesized beam of $\Theta_{\rm maj}=3.537$~mas (white filled circle, bottom-left corners) and aligned to the position of the central binary system (white crosses). Scale bars are shown in the bottom right corner of each panel. The residual root-mean-square (rms) noise levels of the restored maps are 0.40, 0.43, and 0.50~mJy~beam$^{-1}$ at 5, 8.4, and 15~GHz, respectively. Contours start at $3\times{\rm rms}$ and increase by factors of 2.}\label{fig:1998_3freq}
\end{figure}

\begin{figure}[h]
\centering
\includegraphics[width=0.9\textwidth]{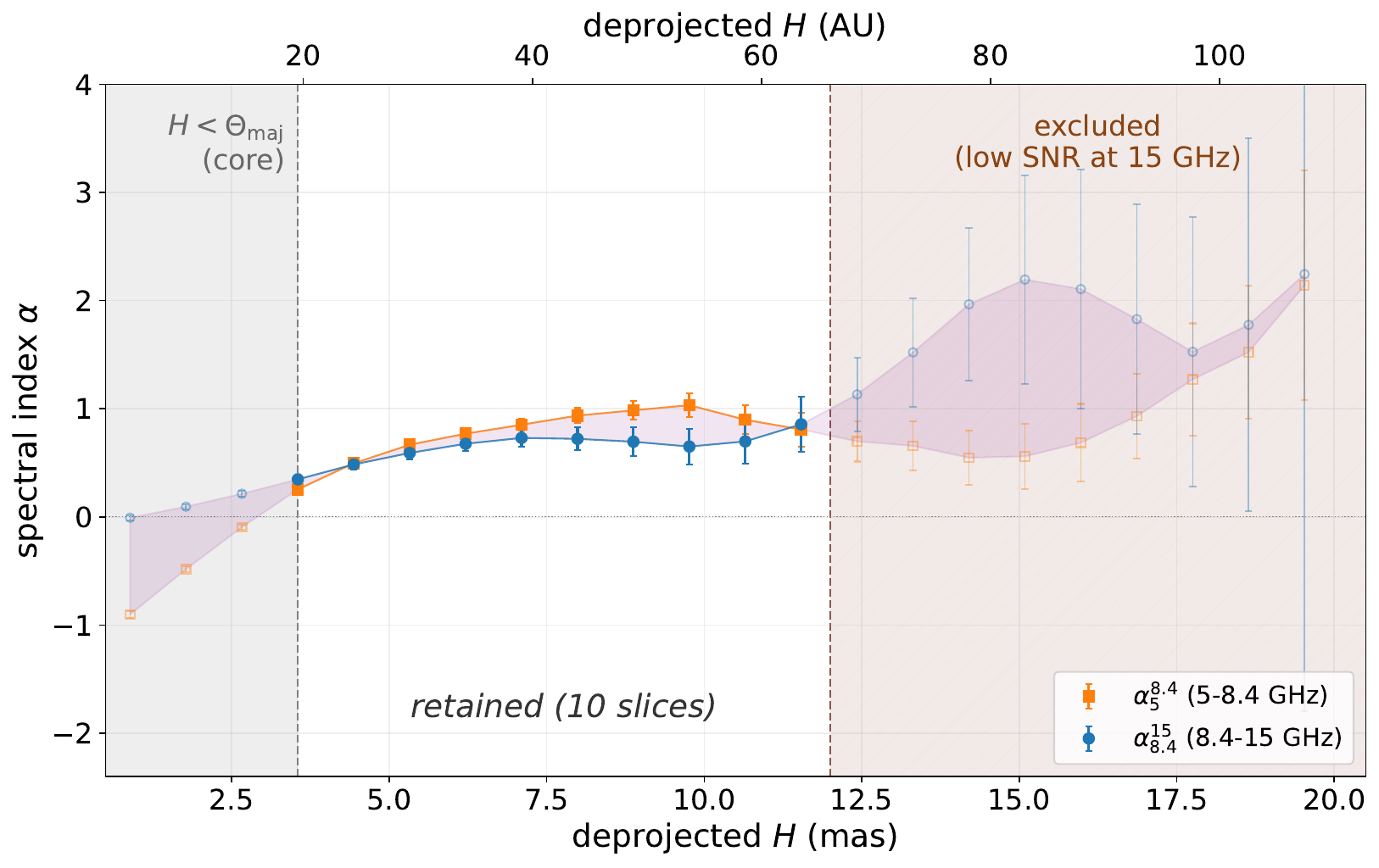}
\caption{Two-point radio spectral indices ($S\propto\nu^{-\alpha}$) versus deprojected distance along the eastern jet of SS~433 on 1998 June 16. Orange squares show $\alpha_{5}^{8.4}$; blue circles show $\alpha_{8.4}^{15}$. The pink shading indicates the separation between the two indices. Open symbols mark slices that are excluded, either because they are contaminated by core free-free absorption ($H<\Theta_{\rm maj}=3.537$\,mas; grey band) or because their 15\,GHz signal-to-noise ratio within the aperture is low ($H\gtrsim 12$\,mas; brown band). Only the ten mid-jet slices shown as filled symbols are retained for the equipartition magnetic-field analysis.}\label{fig:1998_spectral_index}
\end{figure}

\begin{figure}[h]
\centering
\includegraphics[width=0.8\textwidth]{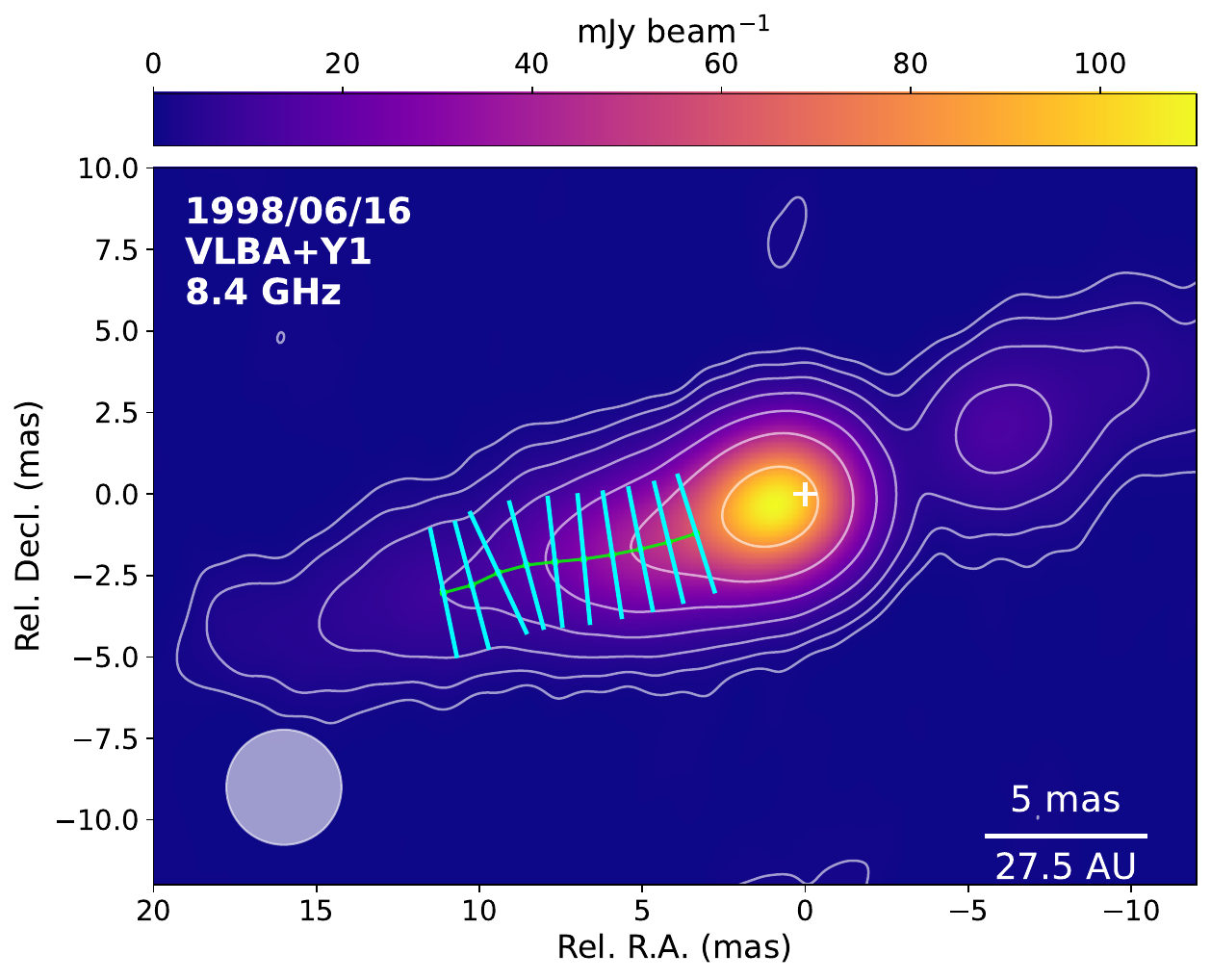}
\caption{The VLBA+Y1 8.4\,GHz image of SS\,433 on 1998 June 16, showing the traced ridge line (green) and the ten transverse slices (cyan bars, with length equal to the fitted transverse FWHM $\Theta_{\rm fit}$). The background contours are drawn at $3\sigma$ times (1, 2, 4, 8, 16, 32, 64, 128), with $\sigma=0.43$~mJy~beam$^{-1}$. The synthesized beam is shown as the white circle in the bottom-left corner, and a scale bar is given in the bottom-right corner. The position of the central binary system is marked by a white cross.}\label{fig:1998_slices}
\end{figure}

\begin{figure}[t]
\centering
\includegraphics[width=1.0\textwidth]{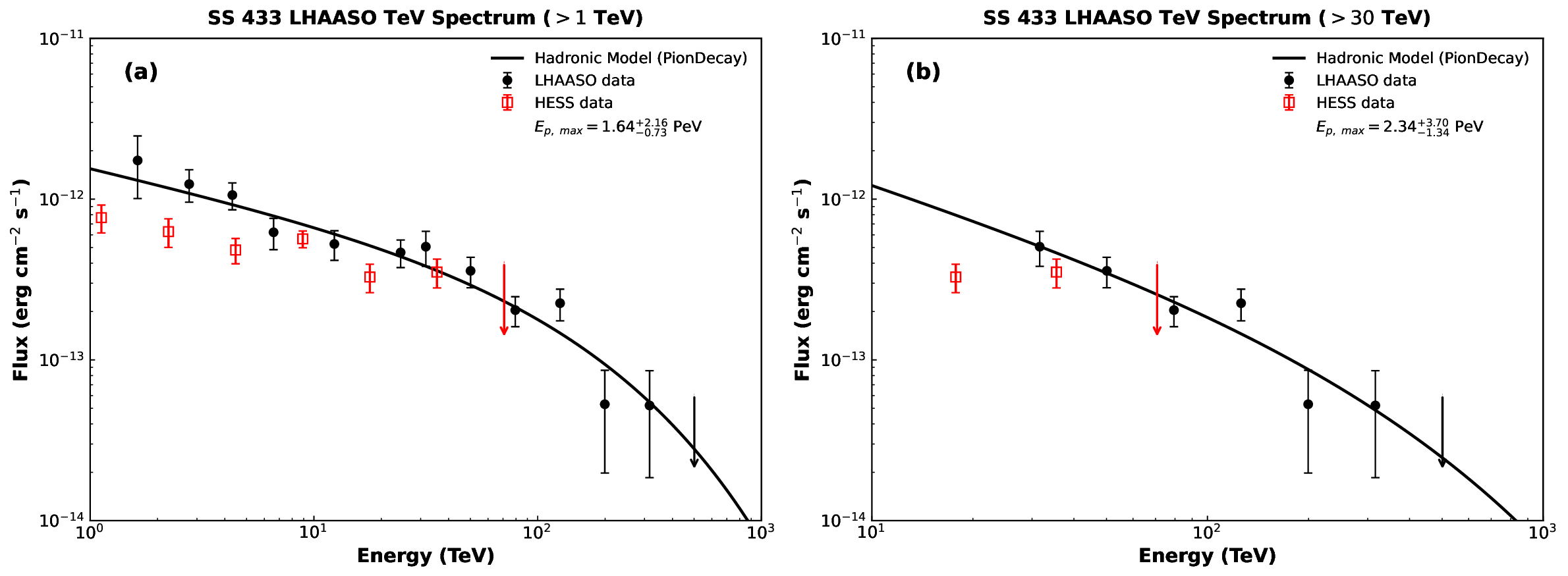}
\caption{Spectral energy distribution of SS~433 fitted with a pure hadronic pion-decay model using LHAASO measurements \citep{LHAASO2025Microquasars}, showing the fit results for two energy ranges ($>1\,\mathrm{TeV}$ and $>30\,\mathrm{TeV}$). The data points below 30\,TeV are from LHAASO-WCDA and each represents the sum of the west- and east-lobe fluxes; all data points above 30\,TeV are from LHAASO-KM2A, with the 30--100\,TeV points representing summed lobe fluxes and those above 100\,TeV originating from a single extended source. The fit to the $>1\,\mathrm{TeV}$ data yields $E_{\rm p,max}=1.64^{+2.16}_{-0.73}\,\mathrm{PeV}$, while the fit to the $>30\,\mathrm{TeV}$ data gives $E_{\rm p,max}=2.34^{+3.70}_{-1.34}\,\mathrm{PeV}$. H.E.S.S.\ measurements \citep{HESS2024SS433} are shown in red for comparison but are excluded from the fit.}
\label{fig:LHAAS0_hadronic_sed}
\end{figure}

\begin{figure}[t]
\centering
\includegraphics[width=0.95\textwidth]{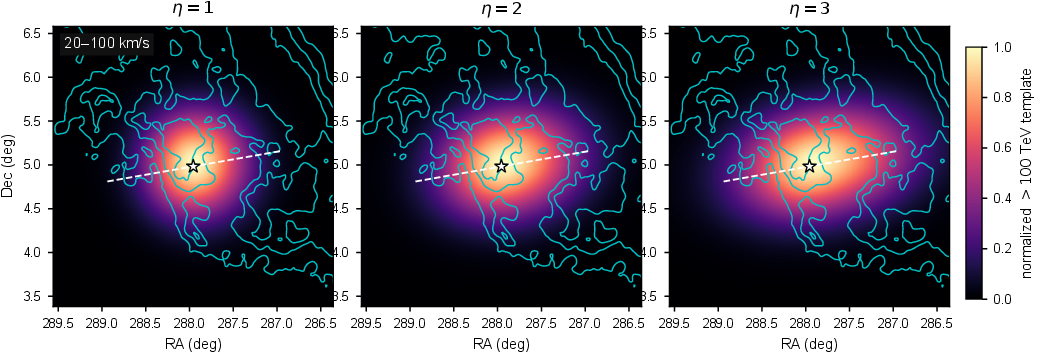}
\caption{Full-window semi-analytic SS~433 $>100\,\mathrm{TeV}$ morphology templates for three values of the transport anisotropy parameter $\eta=D_{\parallel}/D_{\perp}$.
The plotted \(pp\) target map uses the HI4PI $20$--$100\,\mathrm{km\,s^{-1}}$ H\,{\sc i} column-density window, while the cyan contours show the original H\,{\sc i} column-density image used for visual comparison.
The white star marks SS~433 and the dashed white line indicates the adopted jet position angle.
The isotropic case, $\eta=1$, remains nearly round; increasing $\eta$ to 2 and 3 elongates the template along the jet direction, showing why strongly jet-guided transport is less compatible with the approximately compact and weakly elongated LHAASO $>100\,\mathrm{TeV}$ morphology.}
\label{fig:eta_transport_triptych}
\end{figure}

\clearpage

\section*{Supplementary Information}
\subsection*{An illustrative MHD simulation}

We performed an illustrative 2.5D cylindrical
magnetohydrodynamic (MHD) calculation of the SS~433 jet with the
\emph{Message Passing Interface Adaptive Mesh Refinement Versatile Advection
Code} (MPI AMRVAC; \href{https://amrvac.org/}{amrvac.org})~
\citep{Keppens2023AMRVAC}. The calculation adopts axisymmetric cylindrical
geometry, $\partial/\partial\phi=0$, while retaining all three components of
the velocity and magnetic field, and covers $H=30$--$1000\,\mathrm{AU}$
along the jet axis and $r=0$--$180\,\mathrm{AU}$ in cylindrical radius.
One side of the bipolar outflow is simulated with an adiabatic index
$\gamma=5/3$. The jet is injected at $H=30\,\mathrm{AU}$ with
$\beta_j=0.26$ and the fiducial transverse scale
$R_{\rm acc}=\alpha_jH$, where $\alpha_j=0.1$. Its density follows
$\rho_j\propto H^{-2}$ and is normalized to a kinetic power of one jet,
$L_{k,1}=3\times10^{38}\,\mathrm{erg\,s^{-1}}$, giving
$\rho_j(100\,\mathrm{AU})\simeq1.8\times10^{-20}\,
\mathrm{g\,cm^{-3}}$ and
$n_{\rm H}\simeq1.1\times10^{4}\,\mathrm{cm^{-3}}$ for $\mu=1$.
The injected field is predominantly toroidal and follows
\begin{equation}
B_\phi(r,H)=
B_{\rm obs}(H)
\left[
f_{\rm amb}+(1-f_{\rm amb})f_{\rm env}(r)
\right]
\left\{
1-\exp\left[
-\left(\frac{r}{f_{\rm ax}R_{\rm acc}}\right)^2
\right]
\right\},
\end{equation}
where
$B_{\rm obs}(H)=0.19(H/100\,\mathrm{AU})^{-0.50}\,\mathrm{G}$ is the
radio inferred magnetic profile, $f_{\rm amb}=0.01$ sets the residual
ambient field and $f_{\rm ax}=0.35$ regularizes the toroidal component near
the symmetry axis. The thermal pressure is initialized to maintain
approximate transverse total pressure balance, with a reference plasma beta
$\beta_{\rm core}=5$, and the ambient medium is ten times less dense than
the jet and initially nearly stationary. To generate internal shocks and compression structures within the otherwise steady outflow, we imposed weak periodic perturbations in the inlet velocity and density; these perturbations are phenomenological and are not fitted to the observations. The equations are integrated with
the HLLD Riemann solver, a van~Leer limiter, third order strong stability
preserving Runge--Kutta time integration and GLM divergence control, using
adaptive mesh refinement. The flow is evolved to
$t=24\,\mathrm{d}$, slightly longer than the nominal
$\simeq21.6\,\mathrm{d}$ transit time across the domain. Jet precession,
nonthermal particle transport and radiative processes are deliberately
omitted because the calculation is intended as a dynamical consistency test
of the radio inferred magnetic configuration rather than a complete model
of the large scale SS~433 outflow.

As the jet evolves, the imposed perturbations generate propagating compression
structures accompanied by spatial variations in the magnetic field and flow
velocity. We use these evolved quantities to evaluate the local speed weighted
Hillas confinement scale~\citep{Hillas1984},
\begin{equation}
E_{\rm H,\beta}
=
\left(\frac{|\boldsymbol{v}|}{c}\right)
Ze|\boldsymbol{B}|R_{\rm acc},
\qquad
R_{\rm acc}=\alpha_jH .
\label{eq:si_ehbeta}
\end{equation}
Within the jet envelope, $E_{\rm H,\beta}$ spans approximately
$1$ to $7\,{\rm PeV}$ across the simulated region. Extended portions of the
flow reach the $1$ and $2\,{\rm PeV}$ levels, while locally compressed
magnetic structures produce higher rigidity regions. The overall evolution is
consistent with the analytic scaling
$E_{\rm H,\beta}\propto H^{1-q}$ inferred from the radio measurements,
demonstrating that a dynamically evolving jet initialized with the observed
magnetic profile can preserve PeV scale confinement out to
$1000\,{\rm AU}$, corresponding to
$\simeq4.85\times10^{-3}\,{\rm pc}$. Supplementary Fig.~\ref{fig:si_hillas_jet} shows that the magnetic configuration anchored to the radio measurements remains dynamically consistent with PeV proton confinement in the evolved jet: after $24\,{\rm d}$ of evolution, the confinement proxy retains $E_{H,\beta}\gtrsim1\,{\rm PeV}$ throughout the compact baryonic ejecta on the scales covered by the calculation. This provides the confinement condition required to sustain protons at the energies inferred from the gamma-ray observations, thereby supporting the inner jet of SS~433 as a hidden PeVatron.

\setcounter{figure}{0}
\renewcommand{\figurename}{Supplementary Fig.}

\begin{figure}[htbp]
\centering
\includegraphics[width=\textwidth]{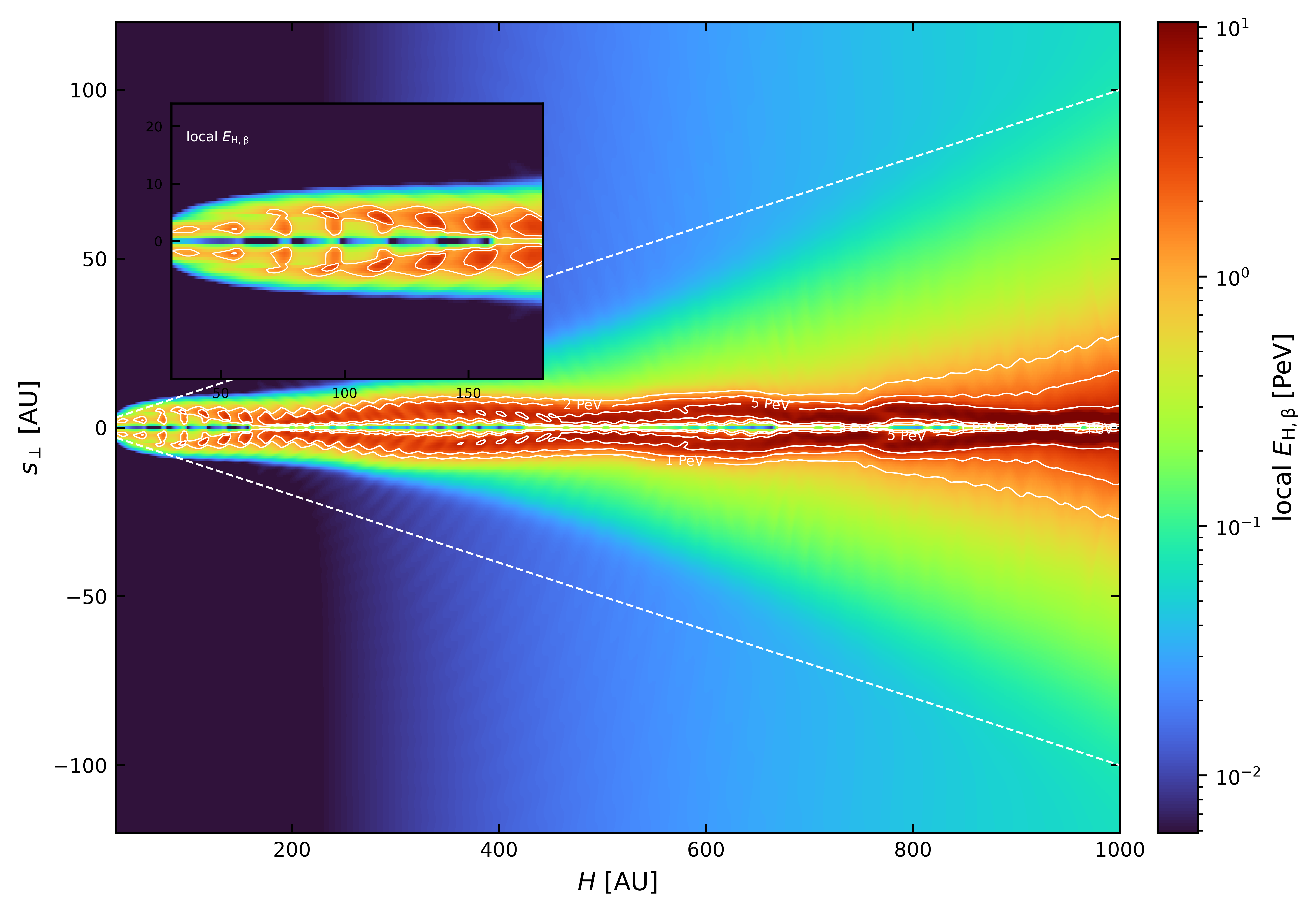}
\caption{Local speed-weighted Hillas confinement proxy $E_{H,\beta}=(|v|/c)\,e\,|B|\,R_{{\rm acc}}$ with $R_{{\rm acc}}=\alpha_jH$
($\alpha_j=0.1$), evaluated on the native fixed-level $384\times2048$ grid at $t=24$\,d. The inflow magnetic field is the radio-derived profile
$B(H)=0.19\,(H/100\,{\rm AU})^{-0.50}$\,G. Contours mark $E_{H,\beta}=1$, $2$, and $5$\,PeV; the inset shows the $30$--$180$\,AU region. The $r\ge0$ cylindrical solution is reflected about the jet axis to display the complete transverse cross-section of the same one-sided jet; it is not a counterjet. Physical jet precession is neither evolved nor displayed in this 2.5-D calculation.}
\label{fig:si_hillas_jet}
\end{figure}

\clearpage
\backmatter

\bibliography{sn-bibliography}

\end{document}